\documentclass[journal]{vgtc}                

\ifpdf
  \pdfoutput=1\relax                   
  \usepackage{graphicx}                
  \DeclareGraphicsExtensions{.pdf,.png,.jpg,.jpeg} 
\else
  \usepackage{graphicx}                
  \DeclareGraphicsExtensions{.eps}     
\fi%

\graphicspath{{figures/}{pictures/}{images/}{./}} 

\usepackage{todonotes}
\usepackage{balance}

\usepackage{microtype}                 
\PassOptionsToPackage{warn}{textcomp}  
\usepackage{textcomp}                  
\usepackage{mathptmx}                  
\usepackage{times}                     
\usepackage{cite}                      
\usepackage{tabu}                      
\usepackage{booktabs}                  
\usepackage{amsmath}
\usepackage{MnSymbol}%
\usepackage[caption=false, font=footnotesize]{subfig}
\usepackage{stmaryrd}
\usepackage{pifont}
\usepackage{wasysym}
\usepackage{multirow}
\usepackage{hyperref}

\newcommand{\ds}[1]{{#1}}

\newcommand{\review}[1]{{\textcolor{black}{#1}}}

\onlineid{1170}

\vgtccategory{Research}
\vgtcpapertype{Algorithm / Technique}

\title{Color Crafting: Automating the Construction \\ of Designer Quality Color Ramps}

\author{Stephen Smart, Keke Wu, Danielle Albers Szafir}
\authorfooter{
\item
Stephen Smart is with the University of Colorado Boulder. E-mail: stephen.smart@colorado.edu.
\item
Keke Wu is with the University of Colorado Boulder. E-mail: keke.wu@colorado.edu.
\item
Danielle Albers Szafir is with the University of Colorado Boulder. E-mail: danielle.szafir@colorado.edu.
}

\shortauthortitle{Smart \MakeLowercase{\textit{et al.}}: Color Crafting: Automating the Construction of Designer Quality Color Ramps}

\abstract{
	Visualizations often encode numeric data using sequential and diverging color ramps. Effective ramps use colors that are sufficiently discriminable, align well with the data, and are aesthetically pleasing. Designers rely on years of experience to create high-quality color ramps. However, it is challenging for novice visualization developers that lack this experience to craft effective ramps as most guidelines for constructing ramps are loosely defined qualitative heuristics that are often difficult to apply. Our goal is to enable visualization developers to readily create effective color encodings using a single seed color. We do this using an algorithmic approach that models designer practices by analyzing patterns in the structure of designer-crafted color ramps. We construct these models from a corpus of 222 expert-designed color ramps, and use the results to automatically generate ramps that mimic designer practices. We evaluate our approach through an empirical study comparing the outputs of our approach with designer-crafted color ramps. Our models produce ramps that support accurate and aesthetically pleasing visualizations at least as well as designer ramps and that outperform conventional mathematical approaches.
} 


\CCScatlist{ 
 \CCScat{K.6.1}{Management of Computing and Information Systems}%
{Project and People Management}{Life Cycle};
 \CCScat{K.7.m}{The Computing Profession}{Miscellaneous}{Ethics}
}

\teaser{
  \centering
  \includegraphics[width=\linewidth]{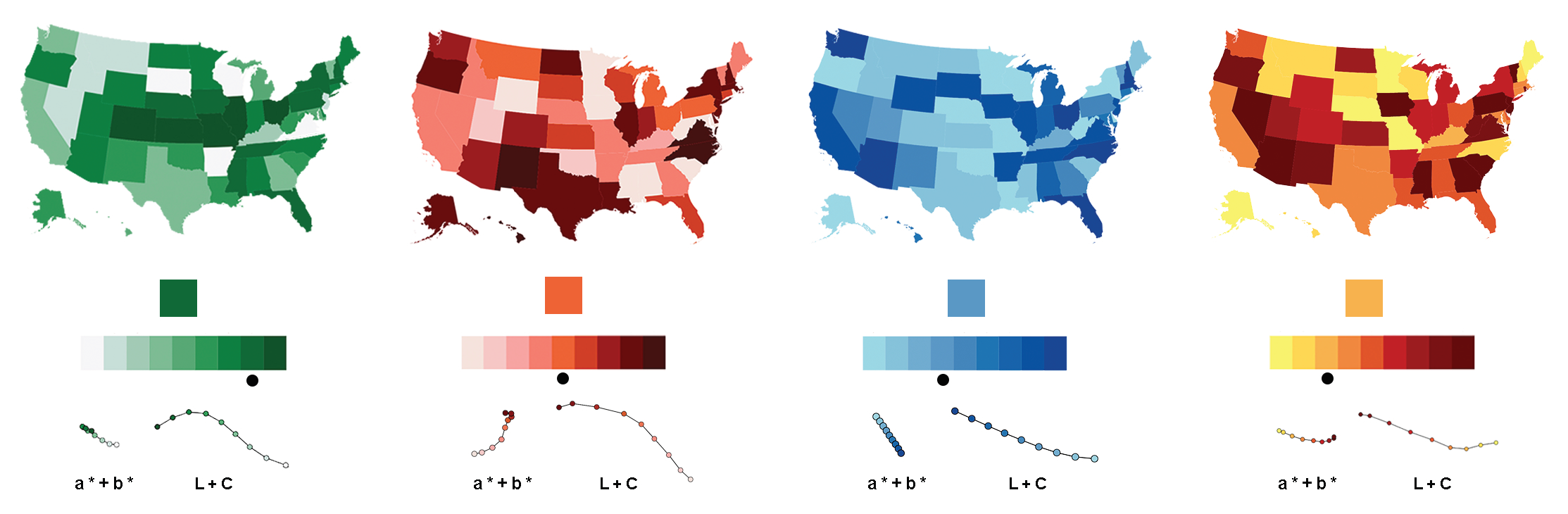}
  \caption{Our approach uses design mining and unsupervised clustering techniques to produce automatically generated color ramps that capture designer practices. The four choropleth maps shown above utilize color ramps generated from our approach. Developers select a single guiding seed color, shown in the squares below each map, to generate a ramp. We then fit curves capturing structural patterns in designer practices in CIELAB (bottom) to these seed colors to generate ramps (middle, seed colors indicated by a black dot). We embody this technique in Color Crafter, a web-based tool that enables designers of all ability levels to generate high-quality custom color ramps.}
	\label{fig:teaser}
}

\vgtcinsertpkg

\begin{document}



\maketitle

\section{Introduction} \label{sec:introduction} 
\ds{Visualizations commonly use \textit{color ramps}
to encode ordered or continuous numeric data.} The specific colors used in a given visualization determine how accurately that visualization communicates the underlying data 
\ds{and influence} subjective impressions of a visualization, such as affect \cite{bartram2017affective}, topical alignment \cite{jahanian2017colors,lin2013selecting}, and aesthetic quality \cite{lau2007towards}. Heuristics for constructing effective encodings have evolved from years of experience by designers. However, these heuristics are often qualitative, ill-defined, and require sufficient expertise to implement. For example, Sloan \& Brown \cite{sloan1979color} recommends that color ramps should use ``a set of colors with an easily remembered order.'' 
While recent efforts mathematically formalize these heuristics into constraints (e.g., Bujack et al. \cite{bujack2018good}), these formalizations still require substantial manual guiding to get from abstract constraints to concrete encodings. Novice visualization designers are currently left with two choices for generating color ramps: to rely on their limited intuitions to craft color ramps of unverified quality or to choose from a small, predefined set of high-quality ramps in tools such as ColorBrewer \cite{harrower2003colorbrewer}.


Existing approaches assert that crafting effective color ramps requires substantial color design expertise and recommend that those lacking this experience should instead draw from preconstructed ramps \cite{wijffelaars2008generating, samsel2018colormoves}. However, this approach restricts the set of available ramps to a small, finite \ds{collection}, limiting the designers' agency and control in creating their visualizations. These limitations are especially prohibitive in situations where specific colors, such as brand colors or semantically meaningful colors \cite{lin2013selecting} are required. Our work aims to give visualization designers of all skill levels an easy way to craft custom designer-quality color ramps. Rather than build up from quantified heuristics, we achieve this goal by modeling designer practices to automatically generate high-quality color ramps from a single seed color. We mine characteristic structures that \ds{a ramp's} sequence of colors traverse in a perceptual color space. These structures capture key variations and patterns that are often overlooked by or difficult to describe using conventional heuristics.

\begin{figure}[t]
	\centering
	\includegraphics[width=\linewidth]{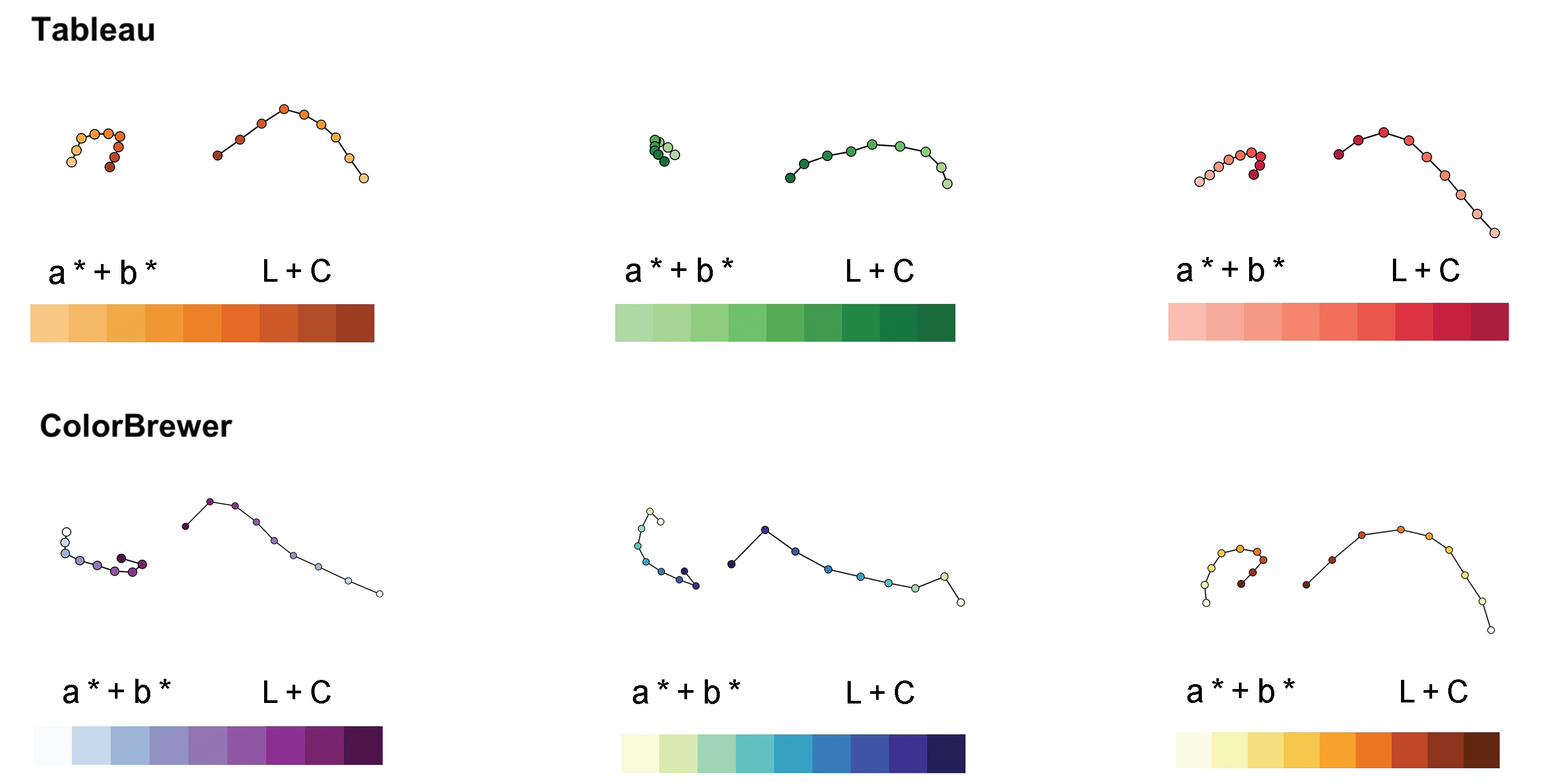}
	\caption{A sample of six single and multihue handcrafted ramps from ColorBrewer and Tableau plotted in CIELAB (projected onto the $a^*-b^*$ and $L-C$ planes) reveals the diversity in structural features found in high quality color ramps.  While many tools linearly interpolate colors in perceptual color spaces to generate ramps, the paths cut by designer ramps traverse non-linear paths, introducing aesthetic variations not well captured by conventional heuristics.}
	\label{fig:designer_curves}
\end{figure}

Color ramps traverse an ordered path through color space that can be modeled as a curve. Common approaches to generating these ramps require designers to select two or more control points and interpolate these control points along either a linear path \cite{mittelstadt2015colorcat,brown} or spline \cite{aisch}. More sophisticated approaches use heuristics from designer practices as constraints applied to interpolating these control points \cite{wijffelaars2008generating,tominski2008task,bergman1995rule}. However, 
these approaches require formalizing a sufficiently complete set of heuristics and rely heavily on the control points to determine the length and direction of the path traversed by the ramp. Our discussions with designers and preliminary analyses of handcrafted ramps reveal features of handcrafted ramps not well-captured by these heuristics (Figure \ref{fig:designer_curves}). Our approach instead builds on the ground truth practices embodied by handcrafted designs to generate a set of curves modeling designer practices, including the direction and structure of colors within the ramp. Designers can seed these curves using a single desired color to generate a collection of ramps reflecting expert practices (Figure \ref{fig:teaser}).

We generate models of common design practices in ramps through design mining: using knowledge discovery techniques to understand design demographics, automate design curation, and support data-driven design tools \cite{kumar2013webzeitgeist}. We assembled and mined a repository of 222 designer-crafted color ramps from popular visualization tools and used these ramps as inputs to two clustering algorithms to construct models that capture common design patterns. Our models represent the continuous paths color encodings traverse in color space in order to capture structural features used by designers
such as the 
twists and kinks in the designer curves shown in Figure \ref{fig:designer_curves}. These continuous paths, modeled as interpolating cubic B-spline curves in CIELAB, can be anchored in color space by seeding each curve with a single guiding color and interpolating around this color to construct a set of ramps. We embed these models in Color Crafter, a tool that allows visualization developers to specify individual seed colors to automatically generate high-quality color ramps. Developers can edit these ramps by freely rotating, translating, and scaling the seeded curve models in CIELAB.

We conducted a crowdsourced experiment with designers to compare ramps generated using our approach to both designer-crafted and linearly interpolated ramps. Our experiment measures how accurately people can identify values using color-coded scatterplots, heatmaps, and choropleth maps as well as the aesthetic value of each visualization. 
Our findings show that our automatically-generated color ramps perform at least as well as designer color ramps at supporting accurate interpretation and positive aesthetics and outperform linear approaches used in common tools. We also show how this approach can readily reproduce common designer ramps (Figure \ref{fig:designer_recreation}) and create high-quality ramps with traditionally ``ugly'' colors (Figure \ref{fig:uglyramps}). \review{The clusters and models from \ds{our} approach open new directions for future work in understanding what makes encodings effective, including a theoretical evaluation of the structures modeled and missed by our approach.}

\noindent \textbf{Contributions}: Our primary contribution is an approach to automate the construction of color ramps that provide high data interpretation accuracy and aesthetic value. Our approach models designer practices by capturing key features of the paths that designer-crafted color ramps traverse through color space. This approach allows novice visualization developers to readily craft high-quality ramps.
We evaluate this approach in a formal experimental evaluation with expert designers and pair of use cases illustrating the utility of our approach.
We embody these results in Color Crafter, an interactive tool for color ramp construction and editing.

\section{Related Work} \label{sec:related_work}
Building effective color ramps requires an understanding of human visual perception as well as the effect of aesthetics in visualization. While the terminology used to describe color encodings
varies across the literature, 
we refer to \emph{colormaps} as any 
set of colors used to encode data. \emph{Color palettes} use discrete colors to encode categorical data while \emph{color ramps} use sequential or diverging colors to encode quantitative data. We survey prior work related to color perception and aesthetics in visualization to inform our work. We also discuss common guidelines and current techniques for constructing color encodings
and recent methods for design mining and automation in visualization.

\begin{figure*}[t]
	\centering
	\includegraphics[width=2\columnwidth]{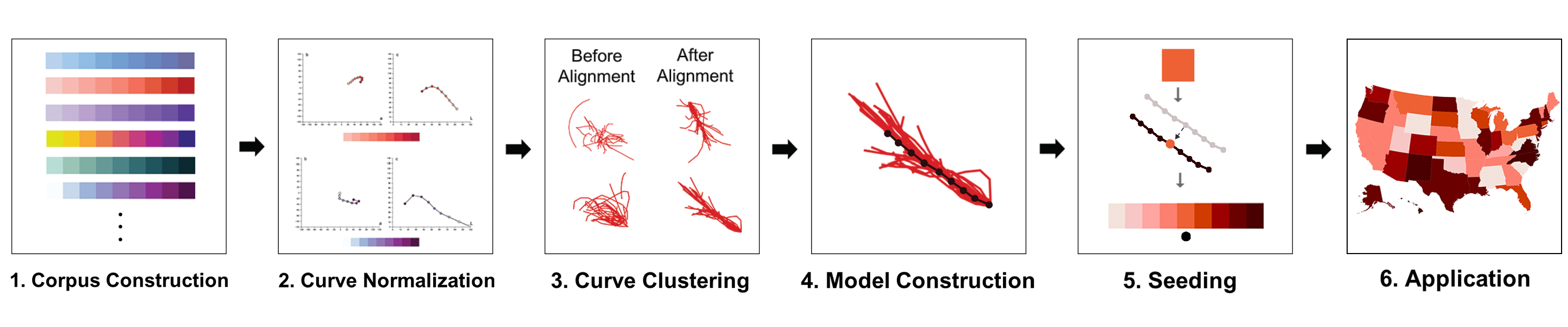}
	\caption{An overview of our algorithm for generating effective color ramps. (1) We collected a corpus of designer-crafted color ramps. (2) We then fit an interpolating cubic B-spline curve to the colors in each ramp in CIELAB, resample each curve to a uniform number of control point colors, and (3) cluster them based on structural patterns in the curves. (4) We then construct a representative model curve for each cluster and (5) use an input seed color (either given by a user or programmatically selected) to anchor the model curve in color space.}
	\label{fig:alg_overview}
\end{figure*}


\subsection{Color Perception and Aesthetics in Visualization}
Color ramps sample colors along a continuous path through a color space. These spaces provide mathematical representations that allow designers to computationally reason about the relationship between colors.
There are artistic color spaces that focus on intuitive parameters of aesthetics (e.g., HSV \cite{ford1998colour, foley1996computer}) and perceptually-based spaces derived from empirical studies of color difference perceptions (e.g., CIELUV and CIELAB \cite{colorimetry1986cie, meyer1980perceptual}). Heer \& Stone \cite{heer2012color} developed models for color naming that enhance visualizations by leveraging the link between visual perception and symbolic cognition. Other models attempt to balance both 
perceptual and artistic approaches (e.g., Munsell \cite{landa2005charting}) or to increase the precision with which they approximate perception (e.g., CIE94 \cite{mcdonald1995cie94} and CIECAM02 \cite{moroney2002ciecam02}). However, these models introduce non-euclidean components that make it difficult to interpolate across. Zeyen et al. \cite{zeyen2018interpolation} present a method for interpolating in non-Euclidean color spaces. While these models correct for imprecisions in color spaces, \ds{their} increased complexity is a trade-off that designers often consider reasonable 
for visualization applications \cite{fairchild2013color}.


Empirical studies of color perception in data visualizations evaluate several different aspects of color's utility (see Kovesi \cite{kovesi2015good} and Silva et al. \cite{silva2011using} for surveys). Several such studies focus on individual features of color perception. For example, Cleveland \& McGill \cite{cleveland1984graphical} found that different color channels communicate data less precisely than other visual channels such as size and position. MacEachren et al. \cite{maceachren2012visual} measure the effectiveness of color and other channels at communicating uncertainty in visualizations. Others studies measure how visual channels such as size \cite{szafir2018modeling} and shape \cite{smart2019measuring} affect color perception.

Other work evaluates full color encodings such as evaluating performance across different color ramps and palettes \cite{liu2018somewhere,bartram2017affective, ware2017evaluating, ware1988color, dasgupta2018effect}. For example, Padilla et al. \cite{padilla2017evaluating} assess trade-offs in binning color ramps, finding that binned ramps often allow analysts to more accurately estimate values in visualizations. Correll et al. \cite{correll2018value} show how manipulating ramp structure can support uncertainty estimation. Schloss et al. \cite{schloss2019mapping} found that the background color in a visualization affects the inferred color mapping between \ds{color ramps and data}, depending on whether the color ramp varies in \ds{apparent} opacity. Liu \& Heer \cite{liu2018somewhere} evaluate how different design parameters, such as color name and perceptual distance, influence encoding accuracy for popular color ramps. These studies provide grounded perceptual insight into aspects of effective color encodings that designers can draw upon when creating encodings.


\ds{However,} leveraging color perception alone is insufficient to create high-quality color ramps. Visual aesthetics 
\ds{significantly} affect factors such as the perceived usability, satisfaction, and pleasure related to a visual representation \cite{moshagen2010facets}.
Several models of aesthetics in visualization have been proposed. These models identify general aspects of aesthetic that are key to positive perceptions of a visualization \cite{lau2007towards, filonik2009measuring, lang2009aesthetics, moshagen2010facets, behrisch2018quality}.
Palmer et al. \cite{palmer2013visual} surveys the aesthetic preferences that have been empirically studied in cognitive science. For example, Western individuals 
\ds{generally} prefer cooler colors such as blue and green over warmer colors such as red and orange \cite{hurlbert2007biological, ling2007new} and more saturated colors compared to less saturated colors \cite{palmer2011ecological}. These effects change according to demographic factors such as age, gender, and culture \cite{palmer2013visual}. Professional designers use knowledge from aesthetic studies to build pleasurable and engaging color encodings. However, models of aesthetic preference focus more on evaluating rather than generating effective visualizations. Our approach allows us to capture aspects of aesthetics functionally embedded into encodings by designers.

\subsection{Guidelines and Techniques}
Approaches to color ramp design combine aspects of color perception and aesthetics. Guidelines about what makes color ramps effective are often derived from designer experience or empirical data (see Bujack et al \cite{bujack2018good} and Zhou \& Hansen \cite{zhou2016survey} for surveys). These guidelines can be either perceptual (e.g., colors should be discriminable) or aesthetic (e.g., colors should be harmonious). 
For example, Brewer \cite{brewer1999color} offers qualitative perceptual and aesthetic considerations for designing effective encodings based on her own practices. Tools such as ColorMeasures \cite{bujack2018good} and colorspace \cite{zeileis2019colorspace} use mathematical abstractions of \ds{such} guidelines to provide quantitative insight into colormap quality.

Perceptual guidelines emphasize ways to make ramps intuitively mirror the underlying data.
Sloan \& Brown \cite{sloan1979color} stress that colors in a ramp should be maximally distinguishable and follow an easily remembered order. 
Wainer \& Francolini \cite{wainer1980empirical} 
\ds{note} that defining \ds{an intuitive} order can be challenging. Trumbo \cite{trumbo1981theory} explains that an effective univariate color ramp should have an ordering in one or more retinal variables and contain sufficient separation between colors.
Levkowitz \& Herman \cite{levkowitz1992color} 
\ds{emphasize a need for \emph{perceptual uniformity:}} the concept that colors should convey the differences between the values they are representing. Zhang \& Montag \cite{zhang2006perceptual} evaluate the performance of colormaps constructed in CIELAB and stress the importance of perceptual uniformity. Mittelstädt et al. \cite{mittelstadt2014revisiting, mittelstadt2015efficient} states that high discriminative power is important for effective color ramps and that highly saturated colors help achieve this \ds{goal}.

Aesthetic guidelines help designers consider how colors might work together to increase the appeal of visualizations. For example, Meier et al. \cite{meier2004interactive} recommend using harmonious colors 
that are not limited to a single planar slice of the HSV color space. Moreland \cite{moreland2009diverging} describes ways to utilize aspects of color such as hue and saturation to create aesthetically pleasing diverging color ramps. Zeileis et al. \cite{zeileis2009escaping} offers advice for choosing colors that are appealing, cooperate with each other, and work in any context. Schloss \& Palmer \cite{schloss2011aesthetic} identify preferable color pairs by decomposing empirical factors in aesthetic preference. Most of these design guidelines are qualitatively defined, involve explicit trade-offs between aesthetics and perception, and require considering the context of a visualization and sophisticated knowledge of color spaces to implement correctly.


In practice, many conventional guidelines, such as a preference towards linearity or uniform differences \cite{levkowitz1996perceptual, levkowitz1992color, pizer1981intensity, tajima1983uniform} may not be followed by experts \cite{brewer1999color}. 
This \ds{deviation} makes it complex for less experienced designers to utilize these guidelines for building effective color ramps. Several tools have been developed to aid visualization developers in using color. Samsel et al. \cite{samsel2015colormaps} leverages the author's artistic experience to craft ramps explicitly for earth science data. ColorBrewer \cite{harrower2003colorbrewer} is a widely-used tool for delivering high-quality color ramps for both categorical and ordered encodings. Models in ColorBrewer were manually crafted using expert knowledge to select and adjust colors over perceptually-ordered spaces. 

\ds{Other tools aim to support visualization developers in crafting color encodings. For example,} Wijffelaars et al. \cite{wijffelaars2008generating} developed a technique to 
\ds{construct} univariate lightness-ordered color ramps as curves parameterized \ds{according to}
qualitative observations of designer ramps. Tominski et al. \cite{tominski2008task} describes a color encoding technique that 
\ds{maximizes effectiveness with respect to specific visualization tasks} such as comparison, localization, and identification.
ColorCAT \cite{mittelstadt2015colorcat} uses constraints regarding analysis tasks and color-vision deficiencies to select colors and linearly interpolates these colors in CIECAM02.
Tools such as Colorgorical \cite{gramazio2017colorgorical}, PRAVDAColor \cite{bergman1995rule}, Tree Colors \cite{tennekes2014tree}, iWantHue \cite{iWantHue}, and VizPalettes \cite{VizPalettes} use rule-based methods to help craft color encodings that adhere to design guidelines. These approaches allow visualization designers to construct effective encodings; however, they generally 
\ds{require} designers to manually specify multiple control points or to tune automatically generated parameters. Generally speaking, increasing the points and parameters necessary to generate an encoding exchanges flexibility for simplicity: designers require more expertise to effectively use the tool. Further, these approaches do not capture nuanced elements of designer practices, such as the features in Figure \ref{fig:designer_curves}. In this work, we hope to simplify the visualization design process for designers of all skill levels using a design-mining approach that models designer practices without the need for a precise translation from design practice to mathematical rules.

\subsection{Automated Design \& Design Mining in Visualization}
Automating the visualization design process 
\ds{computationally applies} design guidelines to generate effective visualizations. Many tools and methods exist for automating or aiding particular aspects of visualization design such as 
computing mappings between data and predefined colormaps \cite{poco2018extracting, lin2013probabilistic, lee2013perceptually},
general perceptual optimization \cite{demiralp2014learning}, chart type selection \cite{wang2018line}, and transfer function design \cite{sereda2006automating}. Other tools and methods focus on fully or partially automating the visualization design process \cite{koop2008viscomplete, el1997addi}. For example, Mackinlay \cite{mackinlay1986automating} developed an application-independent tool that uses
expressiveness and effectiveness \ds{principles} to automatically build effective visualizations.
Wongsuphasawat et al \cite{wongsuphasawat2016towards, wongsuphasawat2016voyager, wongsuphasawat2017voyager} created Voyager, a system that uses both manual and automated chart specification to support data analysis.
Draco \cite{moritz2019formalizing} uses a set of constraints derived from empirical studies in visualizations to formalize visualization design knowledge.

These tools and methods effectively automate several aspects of visualization design. However, some design principles, especially related to aesthetics, are less understood and are difficult to 
\ds{define} mathematically. We can instead use data mining principles to computationally infer these principles from existing artifacts, a concept known as \ds{\emph{design mining}} \cite{kumar2013webzeitgeist}. For example, Jahanian et al. \ds{leverage} design mining \ds{to model color semantics from magazine cover designs \cite{jahanian2017colors} and visual balance from aesthetically pleasing photographs \cite{jahanian2015learning}.}
Samsel et al. \cite{Samsel_Affect} use color schemes from well-known paintings to generate more engaging and expressive scientific visualizations. Our approach builds on these works, utilizing the principles of design mining to infer aspects of color ramp design directly from high-quality examples.

\section{Modeling Designer Practice}
To allow 
visualization developers \ds{of all skill levels} to easily craft high-quality color ramps, we developed an approach that models designer practices by analyzing the general structure of the continuous curves that designer-crafted color ramps traverse through color space. Mathematically, color ramps are an ordered list of points in a \ds{three}-dimensional space. By treating color ramps as continuous curves fit through the individual colors in \ds{a} source ramp, 
we can analyze the relationships between colors and the general structure that designers 
craft when creating high-quality ramps. While prior approaches have used qualitative observations of curve structures in ramps to formulate explicit constraints on encoding design \cite{aisch,wijffelaars2008generating,bergman1995rule}, we instead use a design mining approach to implicitly model designer practices to generate color ramps using techniques from unsupervised machine learning. This approach first computes a set of normalized curves from a corpus of designer ramps. We cluster these curves according to their structural patterns and generate representative curves for each cluster. We anchor these representative curves in color space using a seeding color positioned according to the luminance distribution of the component curves to generate the final ramp. Once seeded, users can optionally edit ramps by applying standard affine transforms (i.e., rotating, translating, \ds{reflecting}, and scaling) to the seeded curves.

Our approach can be described in five major steps (Figure \ref{fig:alg_overview}):
\vspace{-1mm}
\begin{enumerate}
	\item Corpus Construction\vspace{-9pt}
	\item Curve Normalization\vspace{-9pt}
	\item Curve Clustering\vspace{-9pt}
	\item Model Construction\vspace{-9pt}
	\item Seeding
\end{enumerate}

\begin{figure}[t]
	\centering
	\includegraphics[width=0.9 \linewidth]{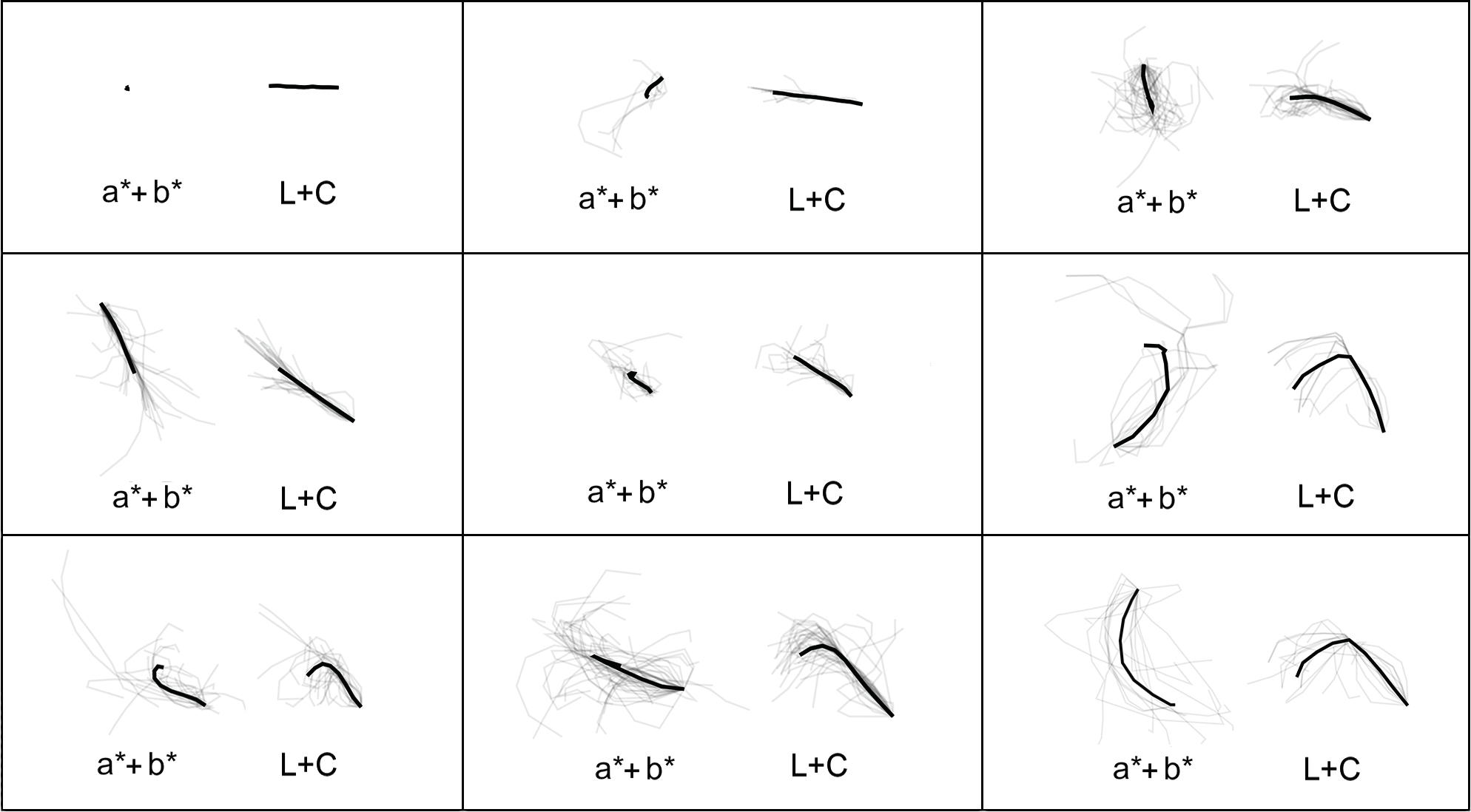}
	\caption{Our algorithm clusters expert-crafted color ramps based on the shapes they traverse in CIELAB (grey), generating nine clusters (projected onto hue and $L^*-C$ planes). We compute representative models for each (black) to capture structural features not well modeled by conventional approaches. \review{Note that curves may be reflected about a central vector during clustering to capture structure agnostic of handedness. We exclude this reflection here to show the structural diversity of the corpus.}  }
	\label{fig:clusters}
\end{figure}

\subsection{Corpus Construction}
Our algorithm uses designer crafted ramps as input data to construct models of effective design. We manually assembled a corpus of 222 unique designer-crafted color ramps from known high-quality sources to provide a ground truth dataset.
This corpus contains 53 ramps from ColorBrewer \cite{harrower2003colorbrewer}, 20 ramps from R,\footnote{\url{https://www.r-project.org/}} 31 ramps from Tableau,\footnote{\url{https://www.tableau.com/}} and 118 ramps from the online designer community ColourLovers\footnote{\url{https://www.colourlovers.com/}}. As ColourLovers is not exclusively for visualization, we only selected color sets that reflected an ordered linear or diverging series of values. This corpus is available at \url{https://tinyurl.com/colorcrafting}.

42 of the 222 were diverging color ramps, while 180 were sequential color ramps. We construct our primary models using the sequential ramps and use additional data from the diverging ramps to provide supplementary parameters for generating diverging ramps (c.f. \S \ref{diverging}).

\subsection{Curve Normalization} \label{subsec:curve_normalization}
The hand-crafted ramps in our corpus contained between 5 and 13 colors. To provide uniform inputs into our clustering algorithms, we resampled these ramps to normalize the number of color points per ramp. We first fit an interpolating cubic B-spline \footnote{\url{https://docs.scipy.org/doc/scipy-0.14.0/reference/generated/\\scipy.interpolate.splprep.html}} through the colors of each ramp in CIELAB. The original ramp colors acted as control points to guide the curve fitting. We then used arc length interpolation to select nine equidistant points along each curve as the colors in our normalized ramps. We chose nine colors as we wish to provide as many control points for our input curves as is reasonable to retain fine-grained features. Designers have previously found nine colors to be a reliable upper bound for the number of perceptually distinct colors in sequential ramps \cite{harrower2003colorbrewer}. We use the normalized curves as inputs to a set of clustering functions to elicit common design patterns across ramps.


\subsection{Curve Clustering}
We observed common design patterns occurring across subsets of ramps, such as characteristic twists in hue space or smaller changes in chroma at the ends of ramps (Figure \ref{fig:designer_curves}). Key features of these patterns are difficult to express mathematically, but are salient in the structures ramps traverse through color space. Our models aim to capture these structures in order to replicate designer practices in constructing color encodings. We cluster the normalized curves computed from our corpus according to different aspects of their shape to group color ramps with similar structures. We compute these clusters agnostic of the specific colors used in order to elicit common relative relationships \emph{between} colors regardless of where those specific colors reside in color space.
We used two different unsupervised clustering techniques: one that uses elastic shape descriptors to match weighted features of the curve structures (Bayesian curve clustering \cite{zhang2015bayesian}) and a second that uses structural features related to known color ramp heuristics (k-means clustering).

\subsubsection{Bayesian Clustering} \label{subsubsec:bayesian_clustering}
Our Bayesian clustering approach leverages the algorithm introduced in Zhang et al. \cite{zhang2015bayesian} which uses an elastic shape metric---the square root velocity function, or $SRVF$ \cite{srivastava2011shape}
---to compute structural relationships between curves. One advantage of 
\ds{this} clustering algorithm is that it infers the number of clusters rather than needing to \ds{specify a number}
\emph{a priori}, allowing natural structural patterns to emerge from the data. The algorithm infers these clusters by computing a posterior distribution on the number of clusters through a Markov Chain Monte Carlo procedure based on the Chinese restaurant process \cite{pitman2006combinatorial}.

While this metric focuses on structural features, one challenge of applying this algorithm to color ramps is that the clustering is designed to be scale invariant (i.e., all curves are scaled to the same length). In color ramps, the length of a curve corresponds with the difference between adjacent colors, a critical component \ds{of} discriminability. To include curve length in the clustering, we adapted the algorithm to use a weighted sum of both the original shape invariant metric and an added term, $L$, reflecting the curve length. $L$ was computed by summing the distance (in absolute $\Delta E$) between each color in the ramp. $f_{SRVF}$ represents the distance metric described in Zhang et al. \cite{zhang2015bayesian} which calculates the difference between $SRVF$ values for a pair of curves, $c_i$ and $c_j$. The resulting algorithm \ds{computes} the distance between any two color ramp curves, $c_i$ and $c_j$, as:\vspace{-6pt}

\[ w \times f_{SRVF}(c_i, c_j) + (1 - w) \times |L(c_i) - L(c_j)| \]\vspace{-6pt}

To find the optimal $w$, we clustered the set of color ramps using ten different values for $w$: 0.0 to 1.0 in 0.1 step increments. For each value of $w$, we evaluated the clustering result by computing the \textit{tightness}---mean summed distance between corresponding \ds{control points}
---of each cluster. To compute the tightness of a cluster, first we aligned all of the curves within each cluster by translating each curve to a common starting point, rotating each curve such that the curves are oriented in the same direction (i.e., the vector passing through each curve's first and middle control color point in the same direction),
and, if needed, reflecting the curve with respect to this aligned vector to minimize the distance between other curves in the cluster. \review{As we cluster based on color-agnostic structures in the curves, this reflection allows us to cluster curves of different handedness \cite{ware1988color}.} Tightness is computed as:\vspace{-6pt}

\[ \frac{1}{n(n-1)} \sum_{i=1}^{n} \sum_{j=1}^{n} \sum_{x=1}^{l} dist(c_i(x), c_{j}(x)) \]\vspace{-3pt}

\noindent where $c_i$ and $c_j$ correspond to ramps in the cluster, $c(x)$ is the current \ds{control point}
color \ds{in the aligned ramp},
$n$ is the number of ramps in the cluster, $l$ is the number of control point colors in a normalized ramp ($l=9$ for our implementation),
and $dist$ computes the Euclidean distance between two corresponding colors. 
After running this clustering algorithm for the ten tested $w$ values, $w = 0.5$ resulted in the tightest clusters. \review{This weight balances length and shape equally, whereas smaller $w$ more heavily considers ramp lengths (correlating with bias towards relative discriminability) and larger favors shape (correlating with bias towards relative aesthetic).} The algorithm generated 9 clusters of between 5 and 42 ramps.

\subsubsection{K-means Clustering} \label{subsubsec:kmeans_clustering}
While the Bayesian clustering approach \ds{leverages} elastic shape descriptors to cluster curves based on their overall structures, many established heuristics \ds{for} color ramps explicitly consider features related to the structure of the ramp. For example, discriminability correlates to the distance between sampled color points along the ramp's curve. To prioritize these structural features, we computed a second set of clusters using k-means clustering, where the features used to compute \ds{clusters} correspond to common design recommendations selected from Bujack et al. \cite{bujack2018good} and qualitative observations of our corpus. This approach pairs implicit and explicit features of ramp design, allowing us to analyze curves based on how designers implement guiding heuristics while measuring cluster quality using our tightness metric captures residual aspects of structural similarity outside of these heuristics.


We computed eight features describing various curve characteristics:

\begin{enumerate}
	\item \textbf{Local Angles:} The angle between each color, which captures local variations in \ds{curve} trajectory \vspace{-6pt}
	\item \textbf{Sum of Angles:} Sum of all angles between adjacent colors, approximating overall trajectory\vspace{-6pt}
	\item \textbf{Local Discriminability:} Distance between each set of adjacent colors in $\Delta E$\vspace{-6pt}
	\item \textbf{Length:} Sum of all distances between adjacent colors in $\Delta E$\vspace{-6pt}
	\item \textbf{Speed:} First derivative of the curve 
	at each color\vspace{-6pt}
	\item \textbf{Acceleration:} Second derivative
	of the curve at each color\vspace{-6pt}
	\item \textbf{Curvature:} Curvature of the curve approximated as $1/r$, where $r$ corresponds to the radius of the sphere of best fit to the 3D curve \vspace{-6pt}
	\item \textbf{Turning Points:} Number of local minima and maxima in the curve. This metric captures small-scale variations and ``kinks'' in the curve not well modeled by conventional heuristics
\end{enumerate}

\begin{figure}[t]
	\centering
	\includegraphics[width=0.9 \linewidth]{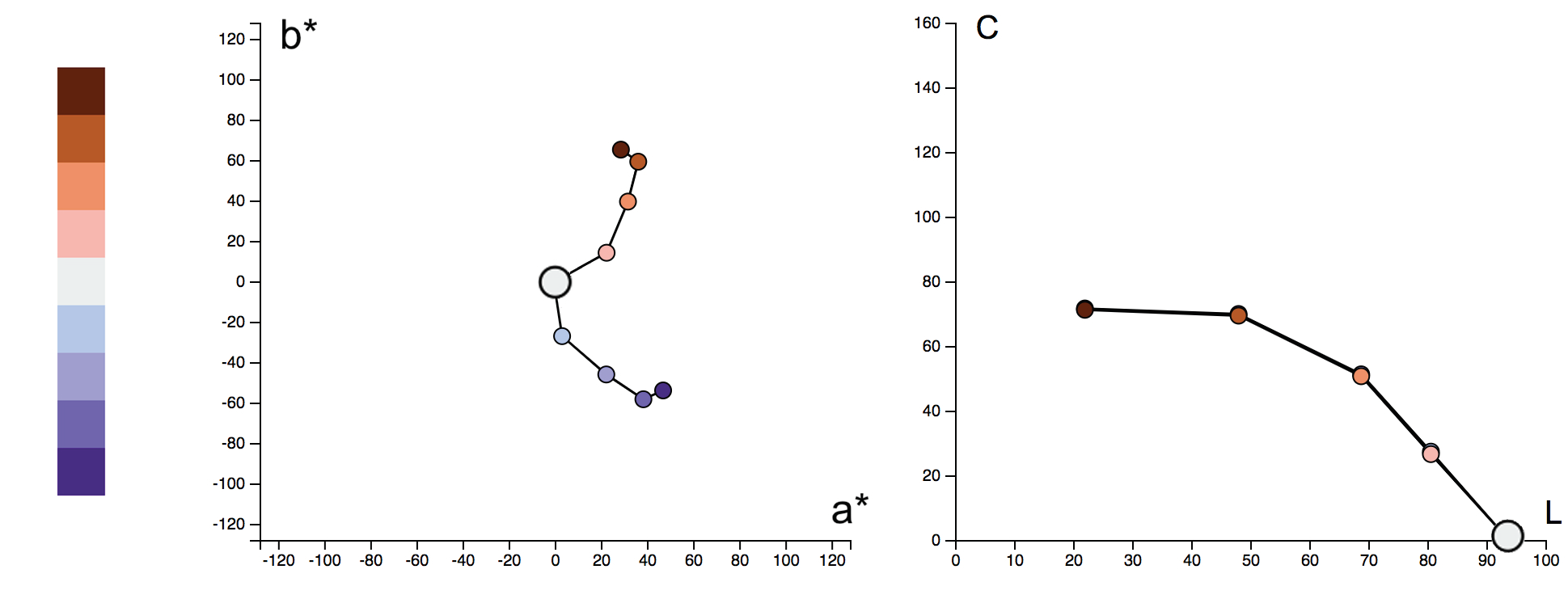}
	\caption{We construct diverging color ramps by appending two sequential model curves together, translating the joining color to a neutral value, and rotating the sequential arms of the color ramp within the range of angles represented in our designer-crafted diverging ramps from our corpus. An example diverging \ds{ramp} crafted using this method is shown above along with the associated curve.}
	\label{fig:divergent}
\end{figure}

\begin{figure}[h]
	\centering
	\begin{tabular}{@{}c@{}}
		\includegraphics[width=.9\linewidth]{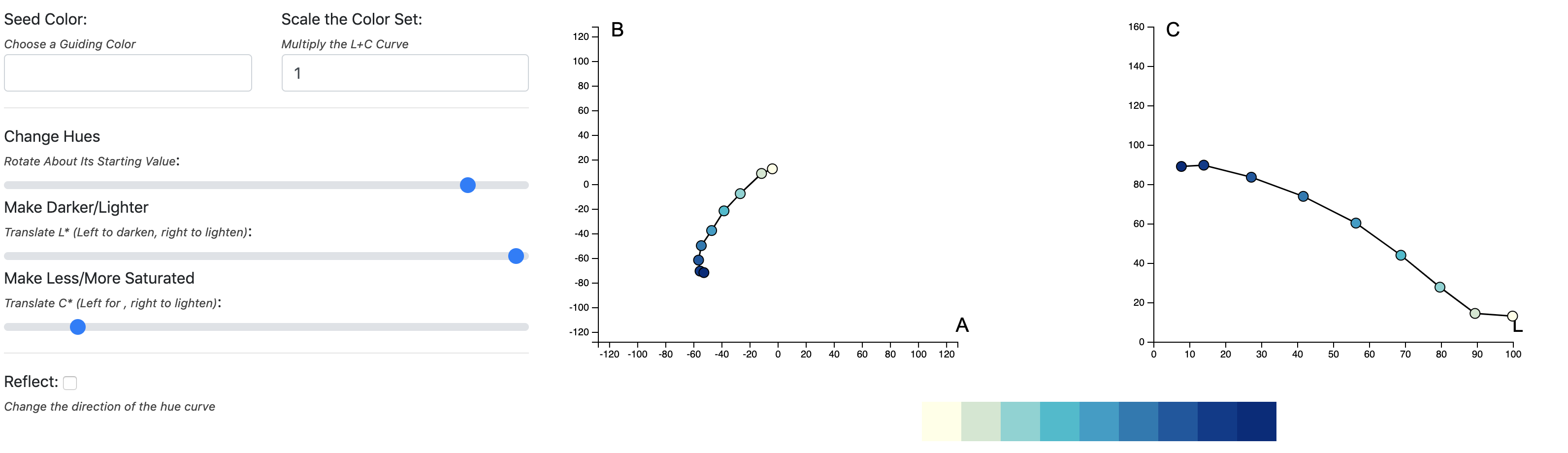} \\[\abovecaptionskip]
		\small (a) Color Crafter tool interface
	\end{tabular}
	\vspace{\floatsep}
	
	\begin{tabular}{@{}c@{}}
		\includegraphics[width=.9\linewidth]{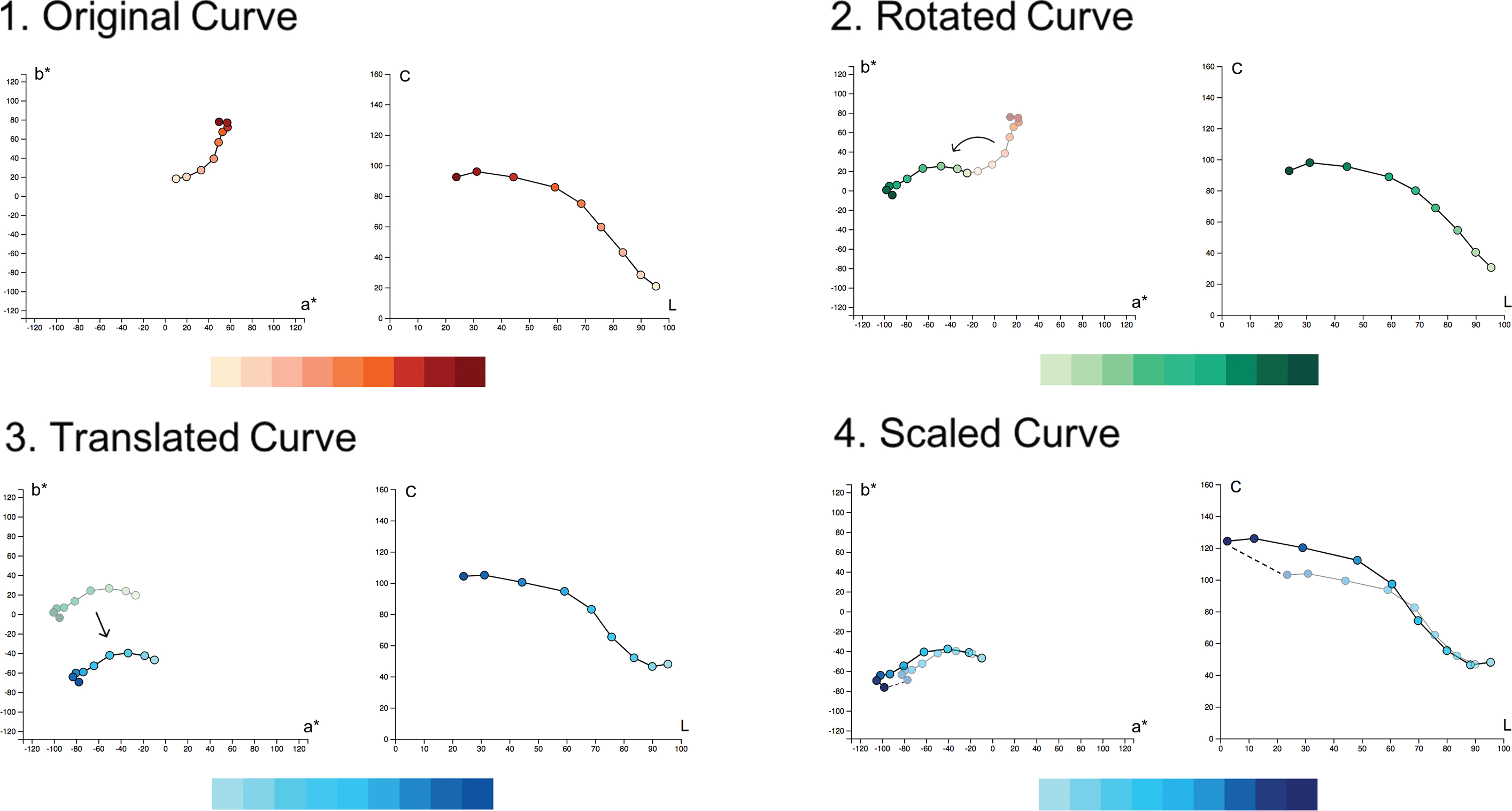} \\[\abovecaptionskip]
		\small (b) Color ramp transformations
	\end{tabular}
	
	\caption{We embody our approach in Color Crafter, a web-based tool for ramp generation and editing. (a) Users can specify target colors and models and (b) to tune these models using affine transforms applied through sliders. These transforms enable users to rapidly refine and explore different encodings while retaining desirable structural properties.}
	\label{fig:curve_editing}
\end{figure}

We then conducted an exhaustive feature selection by computing 255 feature vectors for each curve for $k \in [2, 15]$, \review{with larger $k$ preserving finer differences between models}. We evaluate each cluster configuration using tightness as described in Section \ref{subsubsec:bayesian_clustering}. 
\review{For our corpus,} the ideal configuration of clusters set $k=9$ and used the sum of angles between curve control colors, length, curvature, and number of turning points as features. 

\begin{figure*}[t]
	\centering
	\includegraphics[width=1.75\columnwidth]{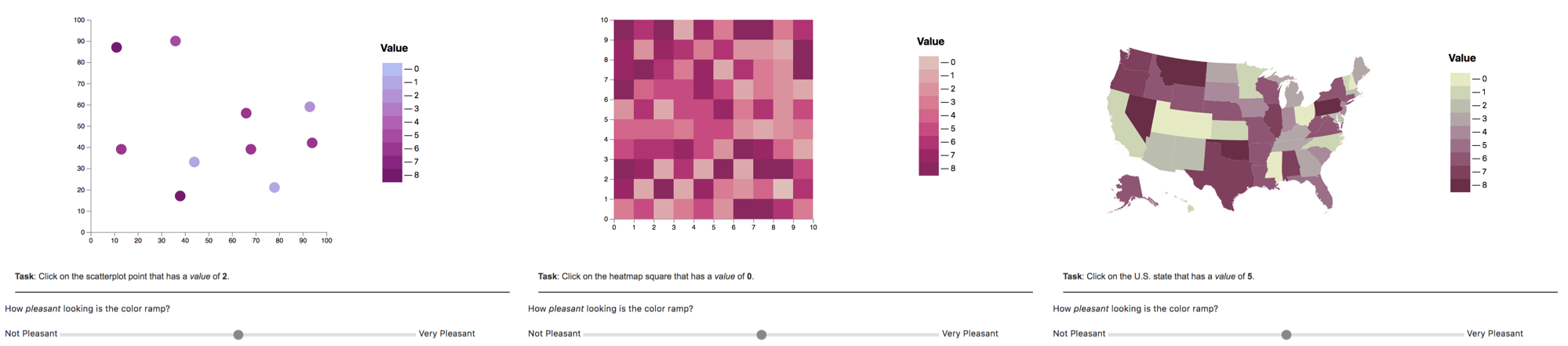}
	\caption{Examples of the study interface used in our empirical evaluation showing each visualization type: scatterplot (left), heatmap (middle), choropleth map (right). Participants clicked on the mark they felt matched a target value and reported how pleasant they found the visualization to be.}
	\label{fig:study}
\end{figure*}

\subsection{Model Construction}
Each cluster produced from the approaches described in Sections \ref{subsubsec:bayesian_clustering} and \ref{subsubsec:kmeans_clustering} captures a specific design pattern found in expert-crafted ramps. 
We model these patterns by constructing a representative curve for each cluster that we can then use to generate ramps reflecting each pattern. We
first align \review{and reflect} the curves within a cluster using the approach described in Section \ref{subsubsec:bayesian_clustering}. 
We then construct a representative curve by computing the mean curve from the set of aligned curves in each cluster. We compute the mean curve as the mean relative position of each control point: \vspace{-9pt}
	
	\[ c'(x) = \frac{\sum_{i=1}^{n} c_i(x)}{n} \]
	\vspace{-9pt}
	
\noindent where $c_i(x)$ is the $x$th color in ramp $i$ and $c'(x)$ is the $x$th control point color in the representative curve. \review{While averaging can compress variation from extremely long or high curvature ramps, characteristic curves tend to qualitatively align well with component curves while making it easier to keep seeded models inside of the gamut. Manual editing can enhance these characteristic curves during ramp design.}

\subsection{Seeding} \label{subsec:seeding}

Aligning the clustered curves before computing the representative curve means that representative curves capture \emph{relative} structures measured in CIELAB: control points reflect the relationship between adjacent colors including the distances and angles between points and the global curve trajectory, but not the actual values of those colors. We can anchor these curves in color space by seeding the model using a single target color. We refer to this color as a \emph{seed color}.

We use a user-specified seed color to translate the representative model curves to a desired region of color space.
As design guidelines heavily emphasize luminance variations in ramp design \cite{brewer1999color}, we first anchor the $L^*$ distribution of the representative curve by translating in $L^*$ to match the average $L^*$ distribution within the corresponding cluster.
We fit a representative curve to a seed color by translating the curve in $L^*$  by $s_L - c_{i,L}(x)$, where $c_{i,L}(x)$ is closest $L^*$ value of a control point color to the seed color. This translation minimizes the amount of displacement from the luminance distribution of the original representative curve. We then translate the curve in the $a^*-b^*$ plane to fit the ramp to the seed color. All subsequent colors in the ramp are computed according to the relative positions of the remaining control points to the seed color.
For example, if the representative curve contains nine colors differing uniformly in $L*$ from 10 to 90 and the seed color is (78, -45, 32), we shift the model curve along $L*$ such that one of the colors contains a lightness of 78, adjusting the $L*$ range to [8,88], with the eighth control color matching the seed color. We then translate the representative curve in the hue plane ($a^*$ and $b^*$) such that the eighth color in the curve is precisely the seed color (78, -45, 32) and calculate the remaining colors from the new coordinates of the curve's control points. Once seeded, developers can rotate, translate, \ds{reflect,} and scale the curve to edit the curve's appearance while retaining the desired structural properties (c.f., \S \ref{subsec:tool}).

\subsection{Constructing Diverging Ramps}
\label{diverging}
Mathematically, diverging color ramps are two sequential ramps joined at a neutral center (or zero) point.  
We can extend sequential models to construct diverging color ramps based on this observation. We \ds{derived} parameters from the set of 42 designer diverging ramps to pair sequential curves to form diverging ramps. We accomplish this by first computing a sequential ramp using the above algorithm, duplicating the ramp's representative curve, and rotating \ds{the duplicated curve} by a controlled amount. As clustering the diverging curves did not provide meaningful clusters, we can instead use statistics describing the angles between sequential arms of designer models to 
\ds{guide} rotations. By default, our approach selects rotations between sequential ramps such that the angle between the curves is equal to 115 degrees\textemdash the mean angle between the sequential arms of the 42 designer-crafted diverging ramps in our corpus. This resulting diverging ramp is then translated in the $a^*-b^*$ plane such that the central point is a shade of gray.

Designers can rotate the arms of the ramp to adjust the hues of the end points and translate the ramp in $L^*$ to adjust the shade of the center point. We constrain hue rotation to between $\pm 60^{\circ}$ to reflect the bounds placed on diverging ramps by designers in our corpus. Figure \ref{fig:divergent} shows an example diverging ramp constructed using this approach.

\subsection{Color Crafter}
\label{subsec:tool}
We embody our approach in Color Crafter, a web-based tool that allows users to seed our representative curves using a specified color, edit that color using a series of affine transformations, and copy the resulting color ramps to the user's clipboard in several common color formats.
The basic interface focuses on providing simple mechanisms for ramp generation (Figure \ref{fig:curve_editing}a). Users input a desired target color and the tool seeds a set of representative curves using that \ds{color}.
Users can edit the curves by translating, rotating, \review{reflecting}, or scaling the component curves (Figure \ref{fig:curve_editing}b). Results of model editing are shown in the curve view alongside a copy of the current ramp. Curves reaching beyond the bounds of the gamut revert to the most recent valid ramp. In future work, we intend to increase the flexibility of the tool to allow expert designers to more freely manipulate the curves and control points.

\section{Evaluation}
We \ds{evaluated} our approach in three ways: 1. an empirical study measuring accuracy and subjective preference with graphic designers, as in Gramazio et al. \cite{gramazio2017colorgorical}; 2. a replication case study showing how our method can readily reproduce ramps echoing popular designer sources, as in Wijffelaars et al. \cite{wijffelaars2008generating}; and 3. a use case evaluation demonstrating the utility of our approach even with conventionally ``ugly'' colors.

\subsection{Empirical Study}

We conducted a 3 (visualization type) x 4 (color ramp type) full factorial within-subjects study drawing on the methodology from Gramazio et al. \cite{gramazio2017colorgorical} to compare our automatically generated sequential color ramps (from both \ds{the} k-means and Bayesian clustering techniques) with linearly interpolated ramps (as in chroma.js \cite{chromajs} and Color Picker for Data \cite{colorpickerfordata}) and designer-crafted ramps. We recruited 35 designers from professional design communities to compare how these 
\ds{ramps} support accurate data interpretation and 
\ds{aesthetic preference.}

\vspace{3pt}
\noindent\textbf{Stimuli:}
Participants were shown a series of visualizations that encoded data using a sequential color ramp and were asked to identify marks in the visualization with a given value and to rate how pleasant they found the visualization. We tested four categories of ramps---designer ramps, ramps constructed using both clustering techniques (k-means and Bayesian), and ramps constructed using linear interpolation in CIELAB---using three different visualizations: a scatterplot, a heatmap, and a choropleth map (Fig. \ref{fig:study}). Scatterplots used 10 16-pixel diameter circular marks rendered at randomly selected x and y coordinates with no overlap between marks. Heatmaps consisted of a \ds{$10 \times 10$} grid where each square was \ds{$30 \times 30$} pixels in size. Choropleth maps were \ds{$550 \times 450$} pixel maps of the United States, \review{providing stimuli testing targets of varying size \cite{stone2014engineering, szafir2018modeling}.} Each mark in a visualization was assigned a value between zero and eight. We visualized these values using one of 108 color ramps (27 per ramp type). We chose nine distinct colors to match the length of our normalized color ramps (c.f., \ds{\S}\ref{subsec:curve_normalization}). Each visualization contained exactly one mark of the target value. All other marks were mapped to a different, randomly selected value.

Our stimuli drew from a \ds{fixed} corpus of 108 color ramps. This corpus contained four sets of 27 ramps reflecting the four categories of ramps used in our experiments: designer, k-means, Bayesian, and linear. Our designer ramps were randomly selected from the corpus of 180 sequential ramps used to guide our models. K-means ramps were generated using our k-means clustering models with 27 unique seeding colors \ds{without manual ramp refinement.}
Bayesian ramps were generated using our unedited Bayesian curves with 27 additional colors. \ds{Linear} ramps linearly interpolated between two control points selected from 27 pairs of pseudorandomly selected colors in CIELAB, reflecting current approaches in many popular tools \cite{chromajs,colorpickerfordata}. We chose seeding colors for these linear ramps that separated the two ends of each linear ramp by a minimum of 40 units in lightness to provide a reasonable comparison to how a developer might craft ramps using this approach \review{(mean $\Delta L^* = 61.3 \pm 9.2$).}

Seeding colors in the k-means, Bayesian, and linear conditions were drawn from a set of 15,581 colors. 
This color set contains all integer colors within $\Delta E=3$ of each color in our \ds{corpus of 180} designer \ds{ramps}
to avoid confounds from hue preferences. 
For example, people tend to find yellow-green tones less appealing overall \cite{gramazio2017colorgorical}, and designers tend to avoid these colors when creating ramps. We sample near the designer colors 
\ds{to help} avoid confounds introduced as a result of the seed color choice: people may rate blue ramps higher overall simply because they prefer blue rather than as a result of the quality of the ramp itself. Given the seed colors are still drawn from a large corpus of possible colors, we anticipate these results 
\ds{reflect} choices novice designers might make, and \ds{we} explore poor color choices in the case study in Section \ref{subsec:uglycolors}. Our method does not explicitly exclude colors as in tools like Colorgorical \ds{as} we want to provide novice designers with sufficient flexibility in crafting ramps according to their needs (e.g., fitting a corporate brand color). Our evaluation instead aims to measure the overall efficiency of our approach in generating ramps that support accurate and aesthetically pleasing visualizations. \review{The distance between ramp endpoints averaged $73.2 \pm 7.8 \Delta E$ for linear ramps, $65.8 \pm 11.9 \Delta E$ for designer ramps, $70.1 \pm 8.6 \Delta E$ for K-Means ramps, and $66.2 \pm 3.0 \Delta E$ for Bayesian ramps.}

\begin{figure}[t]
	\centering
	\includegraphics[width=0.9 \linewidth]{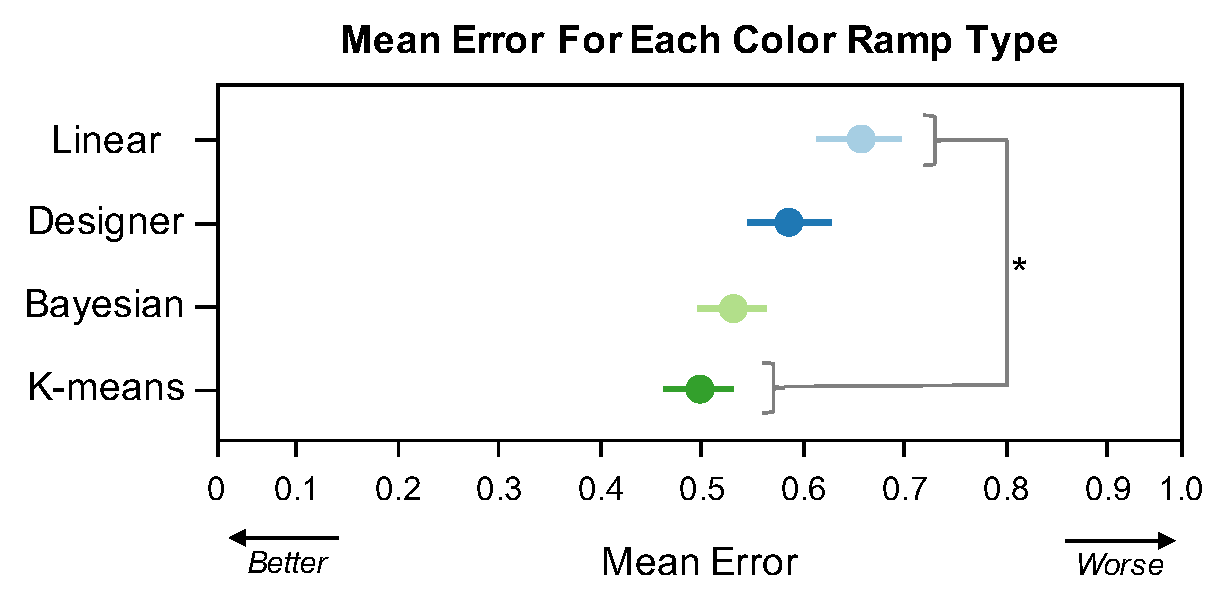}
	\caption{Mean error \ds{(}$|target\textunderscore value - selected\textunderscore value|$\ds{)} for each color ramp type. Error bars indicate 95\% confidence intervals. We found that our participants were significantly more accurate with our ramps than with linear ramps and slightly more accurate than with designer ramps, though the difference was not significant. Significant differences are denoted with asterisks (*: $p < .05$, **: $p < .01$, ***: $p < .001$).}
	\label{fig:results_error}
\end{figure}

\vspace{3pt}
\noindent\textbf{Procedure:}
Our experiment consisted of five phases: (1) informed consent
and screening, (2) preview of visualizations, (3) tutorial, (4) formal study, and (5) demographics questionnaire.
Participants first provided informed consent for their participation and completed a series of digitally rendered Ishihara plates to help screen for color vision deficiencies \cite{clark1924ishihara}. Upon successfully completing the screening, participants then were shown a preview of the full set of 39 visualizations used in the formal study. \review{As noted in prior studies of aesthetics, this preview allows participants to develop a preliminary intuition for the aesthetics of all of the stimuli to help anchor the end-points of their aesthetic preferences prior to beginning the study and to mitigate transfer effects \cite{palmer2013visual,gramazio2017colorgorical}.}
After the preview, participants completed three tutorial questions to clarify any possible ambiguities in the instructions.

In the formal study, each participant completed 39 trials---36 test stimuli plus three engagement checks---presented sequentially in random order to mitigate transfer (e.g., learning or fatigue) effects. Participants clicked on the mark (either a scatterplot point, heatmap square, or U.S. state) that they thought encoded \ds{a} target value. Below the visualization, we included a slider that participants used to report their responses to "How pleasant looking is the color ramp?". 
\ds{The slider} was always initially set to 0. Participants could move the slider to the left\ds{,} which was labeled \textit{Not Pleasant} with a minimum value of -100\ds{,}  or to the right\ds{,}  which was labeled \textit{Very Pleasant} with a maximum value of +100. After each trial, a gray box covered the visualization for 2 seconds to mitigate potential contrast effects between subsequent trials. To ensure honest participation, each experiment contained three trials that were considerably easier to identify the target mark\ds{,} and any participants that failed to correctly answer any of these engagement checks were excluded from our analysis. The full study infrastructure, set of stimulus ramps, and anonymized data is available at \url{https://tinyurl.com/colorcrafting}.

\vspace{3pt}
\noindent\textbf{Participant Recruitment: }
We recruited 35 design practitioners through social media, design interest groups, and online forums. Our participants included designers from the United States, the United Kingdom, India, and China. We excluded four participants from our analysis for either failing to correctly answer the engagement checks (trials that were considerably easier than normal trials), or self-reporting a color vision deficiency or abnormal vision. We analyzed data from the remaining 31 participants ($\mu_{age} = 32.5$ years\ds{;} 13 female, 17 male, 1 DNR), who had an average of 6.2 years of formal design experience.


\begin{figure}[t]
	\centering
	\includegraphics[width=0.9 \linewidth]{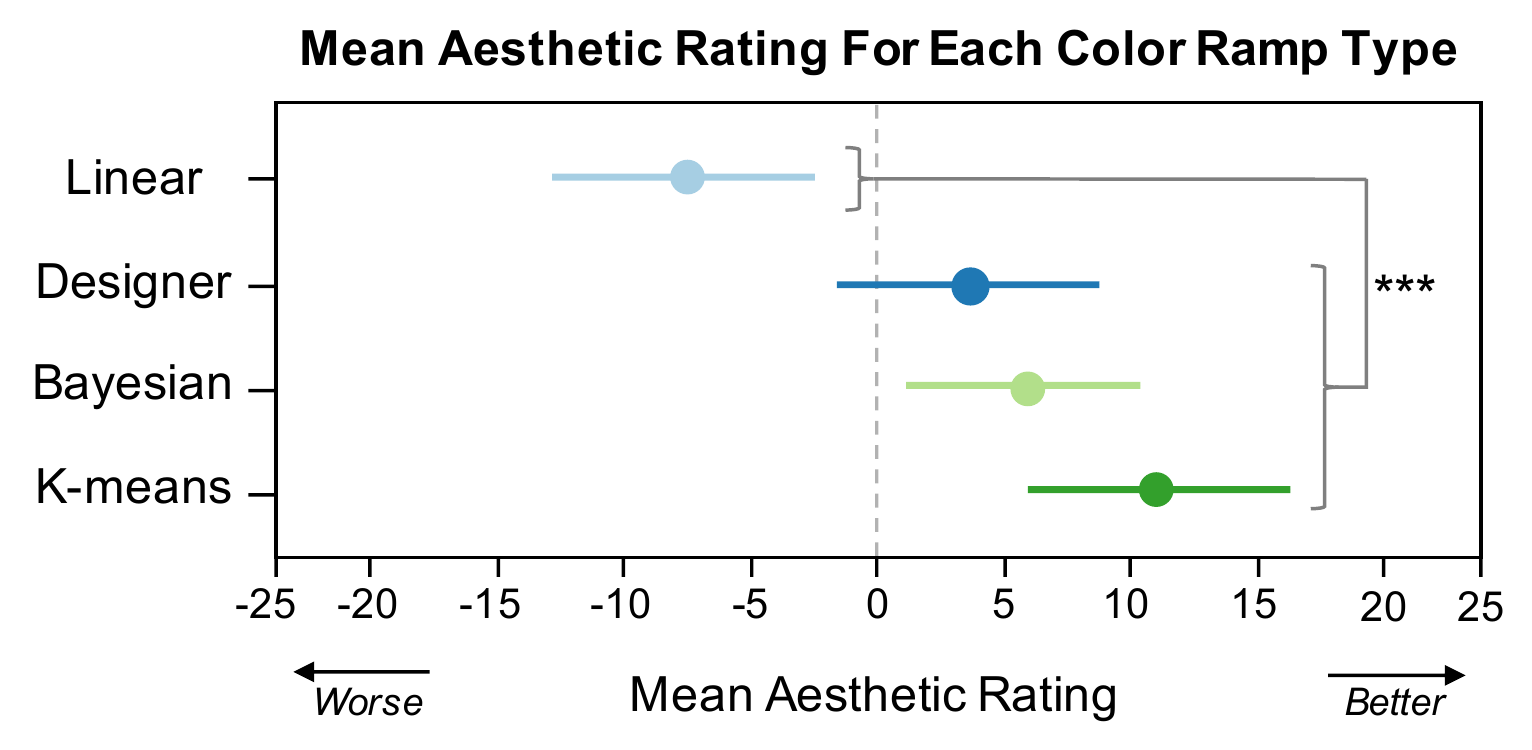}
	\caption{Mean aesthetic rating for each color ramp type. Error bars indicate 95\% confidence intervals. Our automatically generated ramps had slightly higher aesthetic ratings than designer-crafted color ramps (though the difference was not significant) and significantly outperformed color ramps generated using linear interpolation. Significant differences are denoted with asterisks (*: $p < .05$, **: $p < .01$, ***: $p < .001$)}
	\label{fig:results_aesthetic_rating}
\end{figure}

\vspace{3pt}
\noindent\textbf{Results: }
We analyzed task performance and aesthetic ratings using a two-factor repeated measures ANCOVA treating interparticipant variation as a random covariate. All post-hoc analyses used Tukey’s Honest Significant Difference Test (HSD, $\alpha$ = .05). To provide transparency into our effects, we include all data in our supplemental materials and provide both inferential and descriptive statistics in the form of means and 95\% confidence intervals to describe all effects in our study \cite{dragicevic2016fair}. Figures \ref{fig:results_error} and \ref{fig:results_aesthetic_rating} summarize our results. \review{We include the anonymized study data in our OSF supplement for further exploration.}\vspace{1.5mm}

\textit{Error}. We measured error as the absolute value of the difference between the reported and target value.
We found a significant effect of color ramp type on accuracy ($F(3, 90) = 2.835,\ p < .05$). 
While not significant, our generated ramps performed as well or better on average than designer ramps. However, color ramps generated from our k-means clustering technique ($\mu_{error} = 0.495\ \pm\ 0.073$) significantly outperformed linear color ramps ($\mu_{error} = 0.649\ \pm\ 0.088$) at supporting accurate data interpretation (Figure \ref{fig:results_error}).

We also found a significant effect of \ds{visualization} type on accuracy ($F(2, 60) = 5.399,\ p < .01$). Trials involving heatmaps ($\mu_{error} = 0.659\ \pm\ 0.063$) were significantly harder than trials involving scatterplots ($\mu_{error} = 0.519\ \pm\ 0.086$) and choropleth maps ($\mu_{error} = 0.511\ \pm\ 0.058$). \review{Ramps did not perform significantly differently across visualizations}. \vspace{1.5mm}

\textit{Aesthetics}. We measured aesthetic performance as the reported pleasantness of each visualization, with positive values corresponding to positive aesthetics and negative values corresponding to negative aesthetics.
We found a significant effect of color ramp type on aesthetic rating ($F(3, 90) = 8.163,\ p < .001$). Color ramps generated from our k-means ($\mu_{ar} = 9.99\ \pm\ 5.101$) and Bayesian ($\mu_{ar} = 5.48\ \pm\ 4.621$) clustering techniques as well as designer color ramps ($\mu_{ar} = 2.817\ \pm\ 5.358$) were viewed as significantly more aesthetically pleasing than linear ($\mu_{ar} = -7.409\ \pm\ 5.331$) color ramps (Figure \ref{fig:results_aesthetic_rating}).

We also found a significant effect of \ds{visualization} type on aesthetic rating ($F(2, 60) = 7.289,\ p < .001$). Color encodings in heatmaps ($\mu_{ar} = -4.129\ \pm\ 4.443$) were generally perceived as less pleasing than color encodings used in scatterplots ($\mu_{ar} = 5.038\ \pm\ 4.428$) and choropleth maps ($\mu_{ar} = 7.247\ \pm\ 4.444$). \review{While future work should explore these differences, visualizations did not \ds{significantly} affect ramp aesthetics.} 

\subsection{Use Case: Recreating Designer Ramps} \label{subsec:uglycolors}
\begin{figure}[t]
	\centering
	\includegraphics[width=0.75 \linewidth]{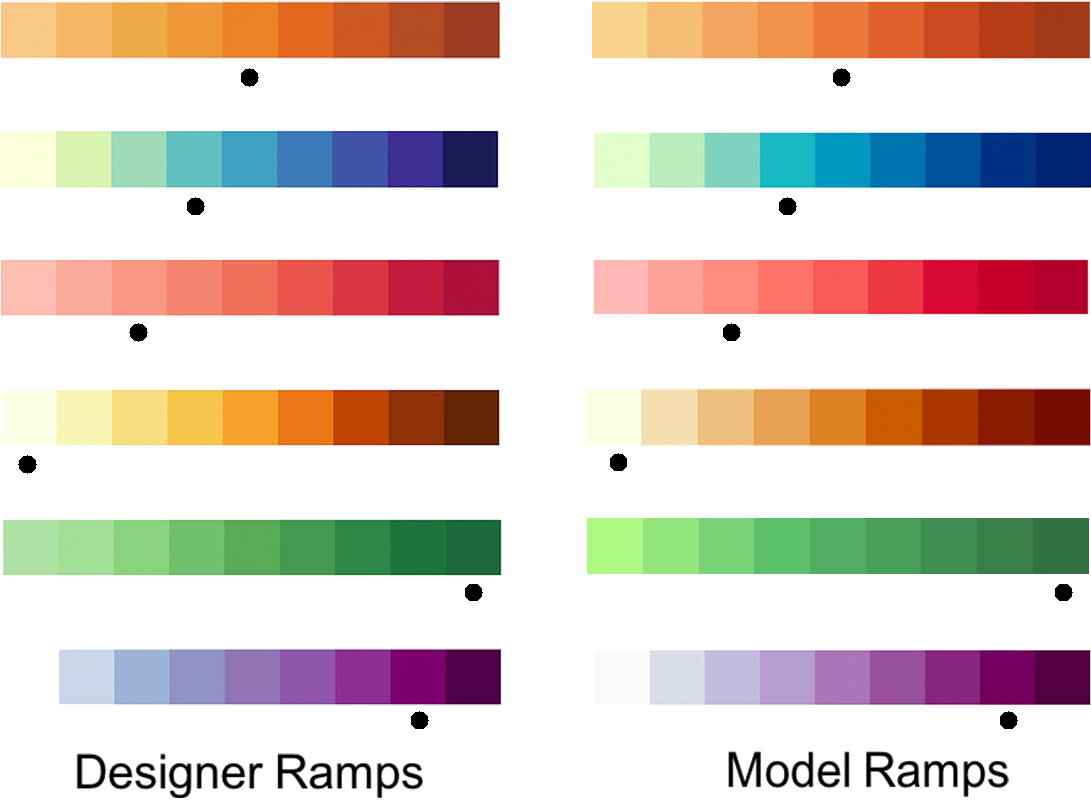}
	\caption{We generated color ramps similar to the 
		designer-crafted ramps in Figure \ref{fig:designer_curves} using seed colors from the original ramps. \ds{While some hue variation is reduced, our} ability to \ds{closely} reproduce these ramps indicates that our models can capture design patterns used by experts.}
	\label{fig:designer_recreation}
\end{figure}

Our clustering approach is intended to capture characteristic design patterns employed in expert-crafted color ramps. However, computing a representative curve may aggregate away fine-scale details present in any individual designer ramp.
To evaluate the effects of this aggregation,
we reconstructed the set of designer-crafted color ramps seen in Figure \ref{fig:designer_curves}. We selected a random seed color from the designer ramp as the input to our algorithm. Using our tool, Color Crafter (Section \ref{subsec:tool}), we applied simple affine transformations to the model curves and selected the model that produced a color ramp most similar to the original designer ramp. A side-by-side comparison of the original designer ramps and the model-generated ramps are shown in Figure \ref{fig:designer_recreation}. While there are some variations from the original ramps, we were able to derive close matches to each of the hand-crafted encodings This replication shows that our models are capable of capturing common design patterns and allows users to readily craft designer-caliber ramps.

\subsection{Use Case: Unpleasant Seed Colors}
\begin{figure}[t]
	\centering
	\includegraphics[width=0.45 \linewidth]{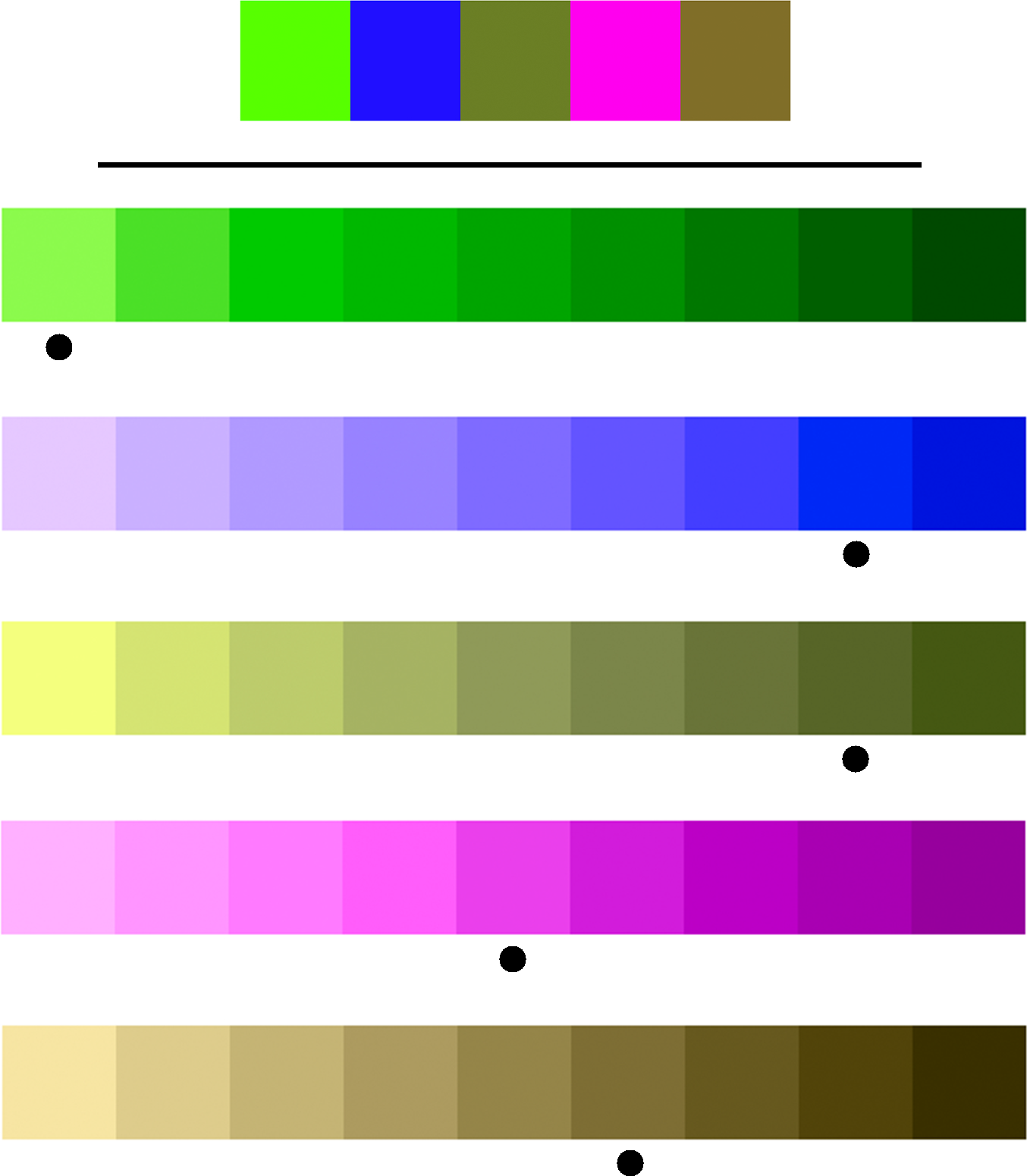}
	\caption{Color ramps generated using ``ugly colors" from the ColourLovers designer community. Seed colors used are shown above and highlighted with a black dot within the ramp. Despite the negative aesthetic of the individual colors, our method generates reasonable encodings even using default structures, suggesting the robustness of this approach. 
	}
	\label{fig:uglyramps}
\end{figure}

While our empirical evaluation focused on a large corpus of designer colors to avoid hue biases, novice designers may select colors with significantly lower aesthetic values to seed ramps. Our method must be robust to poor seed color selection in order to truly support this population. We evaluated this robustness using a worst case design scenario. 
We collected a set of ``ugly'' colors from the ColourLovers design community\footnote{https://www.colourlovers.com/palette/1416250/The\textunderscore Ugliest\textunderscore Colors} to use as seed colors representing actively poor color choices. These ramps are shown in Figure \ref{fig:uglyramps}. The ramps generated using these colors provide aesthetically reasonable color ramps with logical ordering and sufficient discriminability 
even though they use colors conventionally thought of as unpleasant.

\review{Our approach likely creates pleasing encodings from even ``ugly'' individual colors as the model curves prioritize luminance to anchor the generated ramps. As a result, though individual colors are unpleasant, the resulting ramps create aesthetically pleasant combinations of colors---the combinations vary according to designer preferences in lightness and chroma with small, smooth hue adjustments \cite{schloss2011aesthetic}. Colorgorical \cite{gramazio2017colorgorical} uses \ds{this same} idea to estimate pairwise aesthetics. While the colors selected here represent only one community's aesthetic, future studies should evaluate how robust the contrasts in this model are to different seed colors (e.g., yellow-greens \cite{schloss2011aesthetic,gramazio2017colorgorical,palmer2011ecological}).}

\section{Discussion}
Most guidelines for effective color ramp design are qualitative heuristics often originating from years of designer experience and intuition. It is difficult for novice visualization designers to apply these guidelines as they are often ill-defined and require significant expertise to understand and correctly implement. Theoretical formalizations of these guidelines \cite{bujack2018good} help reduce the breadth of knowledge required for them to be effectively applied; however, these formalizations form an incomplete set of constraints better suited to evaluating encodings. Our approach recognizes the practices that designers have developed over many years and leverages a design mining approach to model implicit patterns in these practices to provide visualization designers and developers of all skill levels the ability to easily craft color ramps that support accurate data interpretation and positive aesthetics. Through developing and evaluating our approach, we found:

\begin{itemize}
	\item \textit{Design patterns in color ramps can be quantified using ramp structure}: We can cluster and aggregate the paths traversed by expert color ramps to capture common structural features in design. We computed 18 different patterns to reproduce expert-caliber ramps using individual \ds{seed} colors.\vspace{-3pt}
	\item \textit{Our generated ramps performed at least as well as designer ramps at supporting accuracy}: Color ramps generated using our approach performed at least as well as designer-crafted color ramps at supporting accurate data interpretation. Our k-means clustering technique generated ramps that significantly out-performed linearly interpolated color ramps in a perceptual color space.\vspace{-3pt}
	\item \textit{Our generated ramps performed at least as well as designer ramps at supporting positive aesthetics}: 
	Design practitioners rated ramps from our method as at least as aesthetically pleasing as designer ramps. Both our Bayesian and k-means clustering techniques resulted in ramps that were viewed as significantly more aesthetically pleasing than linearly interpolated color ramps.\vspace{-3pt}
\end{itemize}

\review{Our findings indicate} that the relative structure of the color ramp is the most important factor to generating effective and aesthetically pleasing encodings. While the overall aesthetic quality of our automated ramps may vary, we found that our approach predominantly generated color ramps with pleasing patterns even using random seeding colors. In a pilot study, designers found ramps generated using completely random colors aesthetically comparable to designer ramps ($\mu_{k-means} = 7.548\ \pm\ 6.464,\ \mu_{bayesian} = 8.691\ \pm\ 6.288,\ \mu_{designer} = 12.718\ \pm 6.433$) despite potential confounds from hue preference. These preliminary results coupled with our 
evaluation show that the relationship between colors in a ramp is critical to establishing positive aesthetic. While this finding aligns with prior studies \cite{schloss2011aesthetic}, increasing the number of colors a designer must specify increases the complexity of the design problem, making it harder for those with limited experience to build effective ramps. \review{By capturing the diversity of structures and the nuanced color relationships replicated implicitly in designer encodings, we simplify the design process and empower more people to engage in effective visualization practices.} We anticipate these findings might extend to other encoding design tasks, such as selecting shape configurations. Confirming this hypothesis is important future work.


While we aim to empower novice color designers, our goal is not to replace the designer but rather leverage their expertise to enable novices to produce high-quality color ramps. While explicit, constraint-based methods generate encodings guaranteed to follow good practices, constraint models are often hard to use and miss key, hard-to-define patterns, such as the kinks 
seen in designer-crafted ramps (Figure \ref{fig:designer_curves}) that are critical to effective encoding design. Our approach allows us to detect these features implicitly and provide both novice and skilled designers with a strong and editable baseline. We anticipate that designers could use this approach to quickly generate preliminary encodings and hand-tune those encodings to fit their ideal needs. 

Our approach is most similar to rule-based approaches such as PRAVDAcolor \cite{bergman1995rule} and Colorgorical \cite{gramazio2017colorgorical}. These tools computationally apply perceptual and aesthetic heuristics to derive color encodings for particular tasks. However, most rule-driven tools either require sufficient expertise as the constraint set is incomplete or focus on evaluating rather than generating encodings (e.g., ColorMeasures \cite{bujack2018good} and Pals \cite{zeileis2019colorspace}). Our approach attempts to reduce the expertise needed to craft ramps by minimizing required inputs\ds{: r}amps can be generated using a single color. While our study suggests this approach results in reasonably high quality ramps, our method can generate ramps violating common practices, such as ramps lacking a logical order, but does not do so often. We qualitatively analyzed of the 81 random ramps from our random seed color pilot (\url{https://tinyurl.com/colorcrafting})
and found that while six linear ramps lacked a logical order, our method only generated order-preserving ramps. 

Pairing constraint-based evaluative approaches with our generative model would allow systems to both generate and evaluate encodings at design time. This approach would allow tools to immediately winnow undesirable options from our 18 design patterns before presenting them to users. \review{We could also incorporate additional constraints such as a size-sensitive discriminability \cite{szafir2018modeling} or color-boosting to \ds{mitigate} curvature compression in characteristic curves \ds{(Figure \ref{fig:designer_recreation})}.} Exploring the utility of these generate-and-evaluate paradigms is important future work.

\review{A significant body of research reflects on what makes color ramps effective \cite{zhou2016survey,silva2011using,kovesi2015good}. While our clustering features build on this work, theories around effective perceptual, aesthetic, and cognitive characteristics for color encodings are too numerous to effectively embody in any single tool, yet they are also rife with disagreement and exceptions: ramp design is an art that blends expertise, serendipity, and structured thinking grounded in decades of theory. For example, both designer ramps and our technique often break the convention that a ramp \ds{should include} a starting color with zero chroma (e.g., a gray), instead using yellow or another light hue appropriate to the ramp's theme (Fig. \ref{fig:designer_recreation}). Few ramps in our corpus contain a zero-chroma color. This exemplifies the promise of design-mining methods for generative encodings---our approach breaks conventions in ways similar to designer approaches while preserving common practices such as introducing subtle hue twists. While we focus on optimizing for functionality and design support, our models capture nuanced design structures and offer a lens to explore ramp design theory and heuristics. A full qualitative theoretical analysis of the clusters and structures generated (and not) by this approach is valuable future work but falls outside of the current scope. }


\subsection{Limitations \& Future Work}
The performance of our approach is determined by the quality of the designer-crafted color ramps that we collect. Expanding this corpus may improve the quality of the models being used and provide deeper insight into patterns used in color encoding design. We have released the corpus as an open-source repository (\url{https://tinyurl.com/colorcrafting}) for curated extensibility. 

Our approach allows us to regenerate our models as this corpus grows. \review{However, the success of this technique depends on the models extracted from the corpus: poor quality ramps result in models that reflect poor practices. For example, ramps with high hue variation, such as rainbows, create models with large hue variations and noisy L+C curves \ds{leading} to undesirable color name and luminance variation. In preliminary experiments, such ramps tend to cluster together; however, to generate ramps with unique structures, such as the cubehelix map \footnote{https://jiffyclub.github.io/palettable/cubehelix/}, we may wish to preserve high-variance structures. While our current approaches create ramps with smooth hue, chroma, and luminance variations, ramps with dramatic structural variations would require manual tuning or reweighting of the corpus. Future work should explore how sensitive our approach is to the training ramp distribution. }

As many as 8\% of men and 0.5\% of women suffer from color vision deficiencies \cite{Birch:12}. At present, our approach does not consider CVD as part of the modeling process. Without explicitly considering CVD models as either added constraints to the generation process or as a filter for output ramps, we cannot guarantee that our encodings will be robust for this population. Increasing the robustness of our approach for color vision deficiencies is important future work.

\review{Our evaluation study uses an intentionally simple task to concretely measure perceptual discriminability by how well people can read data values. However, color maps can also be tuned to certain tasks, such as comparison \cite{tominski2008task}, and factors like cognitive load. Future work should further explore how different characteristics of ramp design affect different tasks \ds{to better understand how designers might tailor encodings towards specific goals}. Further, our linear ramps use only two seed colors, whereas many linear interpolation tools allow designers to specify multiple points. We use two points to most closely mimic the single-seed approach used in our technique, but future work could compare against linear interpolation with improved seeding.}

Our approach uses unsupervised clustering methods to identify common structures. However, more sophisticated data mining and generation approaches such as Generalized Adversarial Networks (GANs) \cite{goodfellow2014explaining} may offer alternative methods to clustering. Future work should explore how these methods support effective design automation.


\section{Conclusion}
We present an approach for automatically generating designer-quality color ramps. Our approach utilizes design mining and unsupervised clustering techniques to implicitly learn design patterns from designer color encodings rather than expressing design guidelines as mathematical constraints. We evaluated color ramps generated using our approach through an empirical crowdsourced study with design practitioners. Our results show that ramps generated using our approach performed at least as well as designer-crafted color ramps at supporting accurate data interpretation and positive aesthetics and significantly outperformed color ramps produced using linear interpolation through perceptual color spaces. Our findings also suggest that the relative structure of the curve that a color ramp traverses through color space is more important than the specific colors used for constructing effective color ramps and allows us to capture implicit patterns in expert-caliber visualization design. We hope this work will spark future research on how design mining can empower more people to develop effective visualizations.

\balance

\acknowledgments{
We thank Maureen Stone and the anonymous reviewers for their feedback. This work was supported by
NSF Award \# 1657599.}

\bibliographystyle{abbrv-doi}

\bibliography{template}

\begin{thebibliography}{10}

\bibitem{chromajs}
chroma.js.
\newblock \url{https://github.com/gka/chroma.js/}.

\bibitem{colorpickerfordata}
Color picker for data.
\newblock \url{http://tristen.ca/hcl-picker}.

\bibitem{iWantHue}
iwanthue.
\newblock \url{http://tools.medialab.sciences-po.fr/iwanthue}.

\bibitem{VizPalettes}
Vizpalettes.
\newblock \url{https://www.susielu.com/data-viz/viz-palette}.

\bibitem{aisch}
G.~Aisch.
\newblock Chroma.js color scale helper.

\bibitem{bartram2017affective}
L.~Bartram, A.~Patra, and M.~Stone.
\newblock Affective color in visualization.
\newblock In {\em Proceedings of the 2017 CHI Conference on Human Factors in
  Computing Systems}, pp. 1364--1374. ACM, 2017.

\bibitem{behrisch2018quality}
M.~Behrisch, M.~Blumenschein, N.~W. Kim, L.~Shao, M.~El-Assady, J.~Fuchs,
  D.~Seebacher, A.~Diehl, U.~Brandes, H.~Pfister, et~al.
\newblock Quality metrics for information visualization.
\newblock In {\em Computer Graphics Forum}, vol.~37, pp. 625--662. Wiley Online
  Library, 2018.

\bibitem{bergman1995rule}
L.~D. Bergman, B.~E. Rogowitz, and L.~A. Treinish.
\newblock A rule-based tool for assisting colormap selection.
\newblock In {\em Proceedings of Visualization '95}, pp. 118--125. IEEE, 1995.

\bibitem{Birch:12}
J.~Birch.
\newblock Worldwide prevalence of red-green color deficiency.
\newblock {\em J. Opt. Soc. Am. A}, 29(3):313--320, Mar 2012. doi: {{%
10\hspace{.1pt}\discretionary{.}{%
}{.}\hspace{.4pt}1364\discretionary{/}{%
}{/}JOSAA\hspace{.1pt}\discretionary{.}{%
}{.}\hspace{.4pt}29\hspace{.1pt}\discretionary{.}{%
}{.}\hspace{.4pt}000313}}


\bibitem{brewer1999color}
C.~A. Brewer.
\newblock Color use guidelines for data representation.
\newblock In {\em Proceedings of the Section on Statistical Graphics, American
  Statistical Association}, pp. 55--60. American Statistical Association
  Alexandria, VA, 1999.

\bibitem{brown}
T.~Brown.
\newblock Colorpicker for data.

\bibitem{bujack2018good}
R.~Bujack, T.~L. Turton, F.~Samsel, C.~Ware, D.~H. Rogers, and J.~Ahrens.
\newblock The good, the bad, and the ugly: A theoretical framework for the
  assessment of continuous colormaps.
\newblock {\em IEEE transactions on visualization and computer graphics},
  24(1):923--933, 2018.

\bibitem{colorimetry1986cie}
CIE.
\newblock Cie publication no. 15.2.
\newblock {\em Commission Internationale de leclairage, Vienna}, pp. 19--20,
  1986.

\bibitem{clark1924ishihara}
J.~Clark.
\newblock The ishihara test for color blindness.
\newblock {\em American Journal of Physiological Optics}, 1924.

\bibitem{cleveland1984graphical}
W.~S. Cleveland and R.~McGill.
\newblock Graphical perception: Theory, experimentation, and application to the
  development of graphical methods.
\newblock {\em Journal of the American Statistical Association},
  79(387):531--554, 1984.

\bibitem{correll2018value}
M.~Correll, D.~Moritz, and J.~Heer.
\newblock Value-suppressing uncertainty palettes.
\newblock In {\em Proceedings of the 2018 CHI Conference on Human Factors in
  Computing Systems}, p. 642. ACM, 2018.

\bibitem{dasgupta2018effect}
A.~Dasgupta, J.~Poco, B.~Rogowitz, K.~Han, E.~Bertini, and C.~T. Silva.
\newblock The effect of color scales on climate scientists' objective and
  subjective performance in spatial data analysis tasks.
\newblock {\em IEEE transactions on visualization and computer graphics}, 2018.

\bibitem{demiralp2014learning}
{\c{C}}.~Demiralp, M.~S. Bernstein, and J.~Heer.
\newblock Learning perceptual kernels for visualization design.
\newblock {\em IEEE transactions on visualization and computer graphics},
  20(12):1933--1942, 2014.

\bibitem{dragicevic2016fair}
P.~Dragicevic.
\newblock Fair statistical communication in hci.
\newblock In {\em Modern Statistical Methods for HCI}, pp. 291--330. Springer,
  2016.

\bibitem{el1997addi}
M.~El-Said, G.~Fischer, S.~Gamalel-Din, and M.~Zaki.
\newblock Addi: A tool for automating the design of visual interfaces.
\newblock {\em Computers \& Graphics}, 21(1):79--87, 1997.

\bibitem{fairchild2013color}
M.~D. Fairchild.
\newblock {\em Color Appearance Models}.
\newblock John Wiley \& Sons, 2013.

\bibitem{filonik2009measuring}
D.~Filonik and D.~Baur.
\newblock Measuring aesthetics for information visualization.
\newblock In {\em 2009 13th International Conference Information
  Visualisation}, pp. 579--584. IEEE, 2009.

\bibitem{foley1996computer}
J.~D. Foley, F.~D. Van, A.~Van~Dam, S.~K. Feiner, J.~F. Hughes, J.~HUGHES, and
  E.~ANGEL.
\newblock {\em Computer graphics: principles and practice}, vol. 12110.
\newblock Addison-Wesley Professional, 1996.

\bibitem{ford1998colour}
A.~Ford and A.~Roberts.
\newblock Colour space conversions.
\newblock {\em Westminster University, London}, 1998:1--31, 1998.

\bibitem{goodfellow2014explaining}
I.~J. Goodfellow, J.~Shlens, and C.~Szegedy.
\newblock Explaining and harnessing adversarial examples.
\newblock {\em arXiv preprint arXiv:1412.6572}, 2014.

\bibitem{gramazio2017colorgorical}
C.~C. Gramazio, D.~H. Laidlaw, and K.~B. Schloss.
\newblock Colorgorical: Creating discriminable and preferable color palettes
  for information visualization.
\newblock {\em IEEE transactions on visualization and computer graphics},
  23(1):521--530, 2017.

\bibitem{harrower2003colorbrewer}
M.~Harrower and C.~A. Brewer.
\newblock Colorbrewer. org: an online tool for selecting colour schemes for
  maps.
\newblock {\em The Cartographic Journal}, 40(1):27--37, 2003.

\bibitem{heer2012color}
J.~Heer and M.~Stone.
\newblock Color naming models for color selection, image editing and palette
  design.
\newblock In {\em Proceedings of the SIGCHI Conference on Human Factors in
  Computing Systems}, pp. 1007--1016. ACM, 2012.

\bibitem{hurlbert2007biological}
A.~C. Hurlbert and Y.~Ling.
\newblock Biological components of sex differences in color preference.
\newblock {\em Current Biology}, 17(16):R623--R625, 2007.

\bibitem{jahanian2017colors}
A.~Jahanian, S.~Keshvari, S.~Vishwanathan, and J.~P. Allebach.
\newblock Colors--messengers of concepts: Visual design mining for learning
  color semantics.
\newblock {\em ACM Transactions on Computer-Human Interaction (TOCHI)},
  24(1):2, 2017.

\bibitem{jahanian2015learning}
A.~Jahanian, S.~Vishwanathan, and J.~P. Allebach.
\newblock Learning visual balance from large-scale datasets of aesthetically
  highly rated images.
\newblock In {\em Human Vision and Electronic Imaging XX}, vol. 9394, p.
  93940Y. International Society for Optics and Photonics, 2015.

\bibitem{koop2008viscomplete}
D.~Koop, C.~E. Scheidegger, S.~P. Callahan, J.~Freire, and C.~T. Silva.
\newblock Viscomplete: Automating suggestions for visualization pipelines.
\newblock {\em IEEE Transactions on Visualization and Computer Graphics},
  14(6):1691--1698, 2008.

\bibitem{kovesi2015good}
P.~Kovesi.
\newblock Good colour maps: How to design them.
\newblock {\em arXiv preprint arXiv:1509.03700}, 2015.

\bibitem{kumar2013webzeitgeist}
R.~Kumar, A.~Satyanarayan, C.~Torres, M.~Lim, S.~Ahmad, S.~R. Klemmer, and
  J.~O. Talton.
\newblock Webzeitgeist: design mining the web.
\newblock In {\em Proceedings of the SIGCHI Conference on Human Factors in
  Computing Systems}, pp. 3083--3092. ACM, 2013.

\bibitem{landa2005charting}
E.~R. Landa and M.~D. Fairchild.
\newblock Charting color from the eye of the beholder: A century ago, artist
  albert henry munsell quantified colors based on how they appear to people;
  specializations of his system are still in wide scientific use.
\newblock {\em American Scientist}, 93(5):436--443, 2005.

\bibitem{lang2009aesthetics}
A.~Lang.
\newblock Aesthetics in information visualization.
\newblock {\em Trends in Information Visualization}, 8, 2009.

\bibitem{lau2007towards}
A.~Lau and A.~V. Moere.
\newblock Towards a model of information aesthetics in information
  visualization.
\newblock In {\em 2007 11th International Conference Information Visualization
  (IV'07)}, pp. 87--92. IEEE, 2007.

\bibitem{lee2013perceptually}
S.~Lee, M.~Sips, and H.-P. Seidel.
\newblock Perceptually driven visibility optimization for categorical data
  visualization.
\newblock {\em IEEE Transactions on Visualization and Computer Graphics},
  19(10):1746--1757, 2013.

\bibitem{levkowitz1996perceptual}
H.~Levkowitz.
\newblock Perceptual steps along color scales.
\newblock {\em International Journal of Imaging Systems and Technology},
  7(2):97--101, 1996.

\bibitem{levkowitz1992color}
H.~Levkowitz and G.~T. Herman.
\newblock Color scales for image data.
\newblock {\em IEEE Computer Graphics and Applications}, (1):72--80, 1992.

\bibitem{lin2013selecting}
S.~Lin, J.~Fortuna, C.~Kulkarni, M.~Stone, and J.~Heer.
\newblock Selecting semantically-resonants for data visualization.
\newblock In {\em Computer Graphics Forum}, vol.~32, pp. 401--410. Wiley Online
  Library, 2013.

\bibitem{lin2013probabilistic}
S.~Lin, D.~Ritchie, M.~Fisher, and P.~Hanrahan.
\newblock Probabilistic color-by-numbers: Suggesting pattern colorizations
  using factor graphs.
\newblock {\em ACM Transactions on Graphics (TOG)}, 32(4):37, 2013.

\bibitem{ling2007new}
Y.~Ling and A.~C. Hurlbert.
\newblock A new model for color preference: Universality and individuality.
\newblock In {\em Color and Imaging Conference}, vol. 2007, pp. 8--11. Society
  for Imaging Science and Technology, 2007.

\bibitem{liu2018somewhere}
Y.~Liu and J.~Heer.
\newblock Somewhere over the rainbow: An empirical assessment of quantitative
  colormaps.
\newblock In {\em Proceedings of the 2018 CHI Conference on Human Factors in
  Computing Systems}, p. 598. ACM, 2018.

\bibitem{maceachren2012visual}
A.~M. MacEachren, R.~E. Roth, J.~O'Brien, B.~Li, D.~Swingley, and M.~Gahegan.
\newblock Visual semiotics \& uncertainty visualization: An empirical study.
\newblock {\em IEEE Transactions on Visualization and Computer Graphics},
  18(12):2496--2505, 2012.

\bibitem{mackinlay1986automating}
J.~Mackinlay.
\newblock Automating the design of graphical presentations of relational
  information.
\newblock {\em Acm Transactions On Graphics (Tog)}, 5(2):110--141, 1986.

\bibitem{mcdonald1995cie94}
R.~McDonald and K.~J. Smith.
\newblock Cie94-a new colour-difference formula.
\newblock {\em Journal of the Society of Dyers and Colourists},
  111(12):376--379, 1995.

\bibitem{meier2004interactive}
B.~J. Meier, A.~M. Spalter, and D.~B. Karelitz.
\newblock Interactive color palette tools.
\newblock {\em IEEE Computer Graphics and Applications}, 24(3):64--72, 2004.

\bibitem{meyer1980perceptual}
G.~W. Meyer and D.~P. Greenberg.
\newblock Perceptual color spaces for computer graphics.
\newblock {\em ACM SIGGRAPH Computer Graphics}, 14(3):254--261, 1980.

\bibitem{mittelstadt2014revisiting}
S.~Mittelst{\"a}dt, J.~Bernard, T.~Schreck, M.~Steiger, J.~Kohlhammer, and
  D.~A. Keim.
\newblock Revisiting perceptually optimized color mapping for high-dimensional
  data analysis.
\newblock In {\em EuroVis 2014: the Eurographics Conference on Visualization},
  pp. 91--95, 2014.

\bibitem{mittelstadt2015colorcat}
S.~Mittelst{\"a}dt, D.~J{\"a}ckle, F.~Stoffel, and D.~A. Keim.
\newblock Colorcat: Guided design of colormaps for combined analysis tasks.
\newblock In {\em Eurographics Conference on Visualization (EuroVis)-Short
  Papers. The Eurographics Association}, 2015.

\bibitem{mittelstadt2015efficient}
S.~Mittelst{\"a}dt and D.~A. Keim.
\newblock Efficient contrast effect compensation with personalized perception
  models.
\newblock In {\em Computer Graphics Forum}, vol.~34, pp. 211--220. Wiley Online
  Library, 2015.

\bibitem{moreland2009diverging}
K.~Moreland.
\newblock Diverging color maps for scientific visualization.
\newblock In {\em International Symposium on Visual Computing}, pp. 92--103.
  Springer, 2009.

\bibitem{moritz2019formalizing}
D.~Moritz, C.~Wang, G.~L. Nelson, H.~Lin, A.~M. Smith, B.~Howe, and J.~Heer.
\newblock Formalizing visualization design knowledge as constraints: Actionable
  and extensible models in draco.
\newblock {\em IEEE transactions on visualization and computer graphics},
  25(1):438--448, 2019.

\bibitem{moroney2002ciecam02}
N.~Moroney, M.~D. Fairchild, R.~W. Hunt, C.~Li, M.~R. Luo, and T.~Newman.
\newblock The ciecam02 color appearance model.
\newblock In {\em Color and Imaging Conference}, vol. 2002, pp. 23--27. Society
  for Imaging Science and Technology, 2002.

\bibitem{moshagen2010facets}
M.~Moshagen and M.~T. Thielsch.
\newblock Facets of visual aesthetics.
\newblock {\em International Journal of Human-Computer Studies},
  68(10):689--709, 2010.

\bibitem{padilla2017evaluating}
L.~Padilla, P.~S. Quinan, M.~Meyer, and S.~H. Creem-Regehr.
\newblock Evaluating the impact of binning 2d scalar fields.
\newblock {\em IEEE transactions on visualization and computer graphics},
  23(1):431--440, 2017.

\bibitem{palmer2011ecological}
S.~E. Palmer and K.~B. Schloss.
\newblock Ecological valence and human color preference.
\newblock {\em New Directions in Colour Studies}, pp. 361--76, 2011.

\bibitem{palmer2013visual}
S.~E. Palmer, K.~B. Schloss, and J.~Sammartino.
\newblock Visual aesthetics and human preference.
\newblock {\em Annual Review of Psychology}, 64:77--107, 2013.

\bibitem{pitman2006combinatorial}
J.~Pitman.
\newblock {\em Combinatorial Stochastic Processes: Ecole d'Et{\'e} de
  Probabilit{\'e}s de Saint-Flour XXXII-2002}.
\newblock Springer, 2006.

\bibitem{pizer1981intensity}
S.~M. Pizer.
\newblock Intensity mappings to linearize display devices.
\newblock {\em Computer Graphics and Image Processing}, 17(3):262--268, 1981.

\bibitem{poco2018extracting}
J.~Poco, A.~Mayhua, and J.~Heer.
\newblock Extracting and retargeting color mappings from bitmap images of
  visualizations.
\newblock {\em IEEE transactions on visualization and computer graphics},
  24(1):637--646, 2018.

\bibitem{Samsel_Affect}
F.~Samsel, L.~Bartram, Annie, and Bares.
\newblock Art, affect and color: Creating engaging expressive scientific
  visualization.
\newblock In {\em VISAP: Proceedings of the IEEE VIS Arts Program}, 2018.

\bibitem{samsel2018colormoves}
F.~Samsel, S.~Klaassen, and D.~H. Rogers.
\newblock Colormoves: Real-time interactive colormap construction for
  scientific visualization.
\newblock {\em IEEE computer graphics and applications}, 38(1):20--29, 2018.

\bibitem{samsel2015colormaps}
F.~Samsel, M.~Petersen, T.~Geld, G.~Abram, J.~Wendelberger, and J.~Ahrens.
\newblock Colormaps that improve perception of high-resolution ocean data.
\newblock In {\em Proceedings of the 33rd Annual ACM Conference Extended
  Abstracts on Human Factors in Computing Systems}, pp. 703--710. ACM, 2015.

\bibitem{schloss2019mapping}
K.~B. Schloss, C.~C. Gramazio, A.~T. Silverman, M.~L. Parker, and A.~S. Wang.
\newblock Mapping color to meaning in colormap data visualizations.
\newblock {\em IEEE transactions on visualization and computer graphics},
  25(1):810--819, 2019.

\bibitem{schloss2011aesthetic}
K.~B. Schloss and S.~E. Palmer.
\newblock Aesthetic response to color combinations: preference, harmony, and
  similarity.
\newblock {\em Attention, Perception, \& Psychophysics}, 73(2):551--571, 2011.

\bibitem{sereda2006automating}
P.~Sereda, A.~Vilanova, and F.~A. Gerritsen.
\newblock Automating transfer function design for volume rendering using
  hierarchical clustering of material boundaries.
\newblock In {\em Computer Graphics Forum}, pp. 243--250, 2006.

\bibitem{silva2011using}
S.~Silva, B.~S. Santos, and J.~Madeira.
\newblock Using color in visualization: A survey.
\newblock {\em Computers \& Graphics}, 35(2):320--333, 2011.

\bibitem{sloan1979color}
K.~R. Sloan~Jr and C.~M. Brown.
\newblock Color map techniques.
\newblock {\em Computer Graphics and Image Processing}, 10(4):297--317, 1979.

\bibitem{smart2019measuring}
S.~Smart and D.~A. Szafir.
\newblock Measuring the separability of shape, size, and color in scatterplots.
\newblock In {\em Proceedings of the SIGCHI Conference on Human Factors in
  Computing Systems}. ACM, 2019.

\bibitem{srivastava2011shape}
A.~Srivastava, E.~Klassen, S.~H. Joshi, and I.~H. Jermyn.
\newblock Shape analysis of elastic curves in euclidean spaces.
\newblock {\em IEEE Transactions on Pattern Analysis and Machine Intelligence},
  33(7):1415--1428, 2011.

\bibitem{stone2014engineering}
M.~Stone, D.~A. Szafir, and V.~Setlur.
\newblock An engineering model for color difference as a function of size.
\newblock In {\em Color and Imaging Conference}, vol. 2014, pp. 253--258.
  Society for Imaging Science and Technology, 2014.

\bibitem{szafir2018modeling}
D.~A. Szafir.
\newblock Modeling color difference for visualization design.
\newblock {\em IEEE transactions on visualization and computer graphics},
  24(1):392--401, 2018.

\bibitem{tajima1983uniform}
J.~Tajima.
\newblock Uniform color scale applications to computer graphics.
\newblock {\em Computer Vision, Graphics, and Image Processing},
  21(3):305--325, 1983.

\bibitem{tennekes2014tree}
M.~Tennekes and E.~de~Jonge.
\newblock Tree colors: color schemes for tree-structured data.
\newblock {\em IEEE transactions on visualization and computer graphics},
  20(12):2072--2081, 2014.

\bibitem{tominski2008task}
C.~Tominski, G.~Fuchs, and H.~Schumann.
\newblock Task-driven color coding.
\newblock In {\em 2008 12th International Conference Information
  Visualisation}, pp. 373--380. IEEE, 2008.

\bibitem{trumbo1981theory}
B.~E. Trumbo.
\newblock A theory for coloring bivariate statistical maps.
\newblock {\em The American Statistician}, 35(4):220--226, 1981.

\bibitem{wainer1980empirical}
H.~Wainer and C.~M. Francolini.
\newblock An empirical inquiry concerning human understanding of two-variable
  color maps.
\newblock {\em The American Statistician}, 34(2):81--93, 1980.

\bibitem{wang2018line}
Y.~Wang, F.~Han, L.~Zhu, O.~Deussen, and B.~Chen.
\newblock Line graph or scatter plot? automatic selection of methods for
  visualizing trends in time series.
\newblock {\em IEEE transactions on visualization and computer graphics},
  24(2):1141--1154, 2018.

\bibitem{ware1988color}
C.~Ware.
\newblock Color sequences for univariate maps: Theory, experiments and
  principles.
\newblock {\em IEEE Computer Graphics and Applications}, 8(5):41--49, 1988.

\bibitem{ware2017evaluating}
C.~Ware, T.~L. Turton, F.~Samsel, R.~Bujack, D.~H. Rogers, K.~Lawonn, N.~Smit,
  and D.~Cunningham.
\newblock Evaluating the perceptual uniformity of color sequences for feature
  discrimination.
\newblock In {\em EuroVis Workshop on Reproducibility, Verification, and
  Validation in Visualization (EuroRV3)}. The Eurographics Association, 2017.

\bibitem{wijffelaars2008generating}
M.~Wijffelaars, R.~Vliegen, J.~J. Van~Wijk, and E.-J. Van Der~Linden.
\newblock Generating color palettes using intuitive parameters.
\newblock In {\em Computer Graphics Forum}, vol.~27, pp. 743--750. Wiley Online
  Library, 2008.

\bibitem{wongsuphasawat2016towards}
K.~Wongsuphasawat, D.~Moritz, A.~Anand, J.~Mackinlay, B.~Howe, and J.~Heer.
\newblock Towards a general-purpose query language for visualization
  recommendation.
\newblock In {\em Proceedings of the Workshop on Human-In-the-Loop Data
  Analytics}, p.~4. ACM, 2016.

\bibitem{wongsuphasawat2016voyager}
K.~Wongsuphasawat, D.~Moritz, A.~Anand, J.~Mackinlay, B.~Howe, and J.~Heer.
\newblock Voyager: Exploratory analysis via faceted browsing of visualization
  recommendations.
\newblock {\em IEEE transactions on visualization and computer graphics},
  22(1):649--658, 2016.

\bibitem{wongsuphasawat2017voyager}
K.~Wongsuphasawat, Z.~Qu, D.~Moritz, R.~Chang, F.~Ouk, A.~Anand, J.~Mackinlay,
  B.~Howe, and J.~Heer.
\newblock Voyager 2: Augmenting visual analysis with partial view
  specifications.
\newblock In {\em Proceedings of the 2017 CHI Conference on Human Factors in
  Computing Systems}, pp. 2648--2659. ACM, 2017.

\bibitem{zeileis2019colorspace}
A.~Zeileis, J.~C. Fisher, K.~Hornik, R.~Ihaka, C.~D. McWhite, P.~Murrell,
  R.~Stauffer, and C.~O. Wilke.
\newblock colorspace: A toolbox for manipulating and assessing colors and
  palettes.
\newblock {\em arXiv preprint arXiv:1903.06490}, 2019.

\bibitem{zeileis2009escaping}
A.~Zeileis, K.~Hornik, and P.~Murrell.
\newblock Escaping rgbland: Selecting colors for statistical graphics.
\newblock {\em Computational Statistics \& Data Analysis}, 53(9):3259--3270,
  2009.

\bibitem{zeyen2018interpolation}
M.~Zeyen, T.~Post, a.~A. Hans~Hagen, D.~Rogers, and R.~Bujack.
\newblock {Color Interpolation for Non-Euclidean Color Spaces}.
\newblock In {\em IEEE Scientific Visualization Conference (SciVis) Short
  Papers}. IEEE, 2018.

\bibitem{zhang2006perceptual}
H.~Zhang and E.~Montag.
\newblock Perceptual color scales for univariate and bivariate data display.
\newblock {\em The Society for Imaging Science and Technology (IS\&T)}, 2006.

\bibitem{zhang2015bayesian}
Z.~Zhang, D.~Pati, and A.~Srivastava.
\newblock Bayesian clustering of shapes of curves.
\newblock {\em Journal of Statistical Planning and Inference}, 166:171--186,
  2015.

\bibitem{zhou2016survey}
L.~Zhou and C.~D. Hansen.
\newblock A survey of colormaps in visualization.
\newblock {\em IEEE transactions on visualization and computer graphics},
  22(8):2051--2069, 2016.

\end{thebibliography}
\end{document}